\documentclass[aps,prb,twocolumn,superscriptaddress,floatfix,amssymb,longbibliography]{revtex4-1}

\usepackage{stix}
\usepackage{graphicx}
\usepackage{dcolumn}
\usepackage{amsmath}
\usepackage{amssymb}
\usepackage{mathtools}
\usepackage[usenames, dvipsnames]{color}
\usepackage[pdftex]{hyperref}
\hypersetup{colorlinks=true,linkcolor=blue,citecolor=blue,urlcolor=blue}

\begin{document}

\title{Quasi-2D magnetic correlations in Ni$_2$P$_2$S$_6$ probed by ${}^{31}$P NMR}

\author{A. P. Dioguardi} 
\affiliation{IFW Dresden, Institute for Solid State Research, P.O. Box 270116, D-01171 Dresden, Germany}

\author{S. Selter} 
\author{S. Aswartham} 
\author{M.-I. Sturza} 
\affiliation{IFW Dresden, Institute for Solid State Research, P.O. Box 270116, D-01171 Dresden, Germany}

\author{R. Murugesan} 
\author{M. S. Eldeeb} 
\author{L. Hozoi} 
\affiliation{IFW Dresden, Institute for Theoretical Solid State Physics, P.O. Box 270116, D-01171 Dresden, Germany}

\author{B. B{\"{u}}chner} 
\affiliation{IFW Dresden, Institute for Solid State Research, P.O. Box 270116, D-01171 Dresden, Germany}
\affiliation{Institute for Solid State Physics, Dresden Technical University, TU-Dresden, 01062 Dresden, Germany}
\author{H.-J. Grafe} 
\affiliation{IFW Dresden, Institute for Solid State Research, P.O. Box 270116, D-01171 Dresden, Germany}

\date{\today}

\begin{abstract}
Detailed ${}^{31}$P nuclear magnetic resonance (NMR) measurements are presented on well-characterized single crystals of antiferromagnetic van der Waals Ni$_2$P$_2$S$_6$. An anomalous breakdown is observed in the proportionality of the NMR shift $K$ with the bulk susceptibility $\chi$. This so-called $K$--$\chi$ anomaly occurs in close proximity to the broad peak in $\chi(T)$, thereby implying a connection to quasi-2D magnetic correlations known to be responsible for this maximum. Quantum chemistry calculations show that crystal field energy level depopulation effects cannot be responsible for the $K$--$\chi$ anomaly. Appreciable in-plane transferred hyperfine coupling is observed, which is consistent with the proposed Ni--S--Ni super- and Ni--S--S--Ni super-super-exchange coupling mechanisms. Magnetization and spin--lattice relaxation rate ($T_1^{-1}$) measurements indicate little to no magnetic field dependence of the N{\'e}el temperature. Finally, $T_1^{-1}(T)$ evidences relaxation driven by three-magnon scattering in the antiferromagnetic state.
\end{abstract}

\maketitle

\section{Introduction}
\label{sec:31P_NMR}

Quasi-two-dimensional (quasi-2D) layered van der Waals materials have attracted interest for decades due to the rich variety of magnetic properties and strong electronic correlations~\cite{Le_Flem_1982_MPX3_magn_inter, Brec_1986_MPS3_review, Balkanski_1987_MPX3_spin_order, Grasso_2002_MPX3_family, Manzeli_2017_2D_TMDCs, Kim_2018_NiPS3_correlations, Zeisner_2019_Cr2Ge2Te6_ESR, Zhang_2019_CrSiTe3_correlated_FM} as well as applications in the fields of Li-based energy storage~\cite{Brec_1979_MPX3_Li_intercalation, Grasso_2002_MPX3_family, Jung_2016_Li_intercalation}, optoelectronics and photonics~\cite{Mak_2016_TMDCs_photo_optoelec}, and spintronics~\cite{Zhong_2017_vdW_heterostr}, among other possible next-generation applications~\cite{Manzeli_2017_2D_TMDCs}. It has been recently shown that the magnetism in van der Waals materials persists down to the monolayer limit~\cite{Lee_2016_FePS3_atomic}, devices such as field effect transistors have been demonstrated~\cite{Jenjeti_2018_NiPS3_FET}, and additionally heterostructures can be engineered to explore fundamental physics and produce novel devices for spintronics applications~\cite{Zhong_2017_vdW_heterostr}.

\begin{figure*}
    \includegraphics[trim=0cm 0cm 0cm 0cm, clip=true, width=\linewidth]{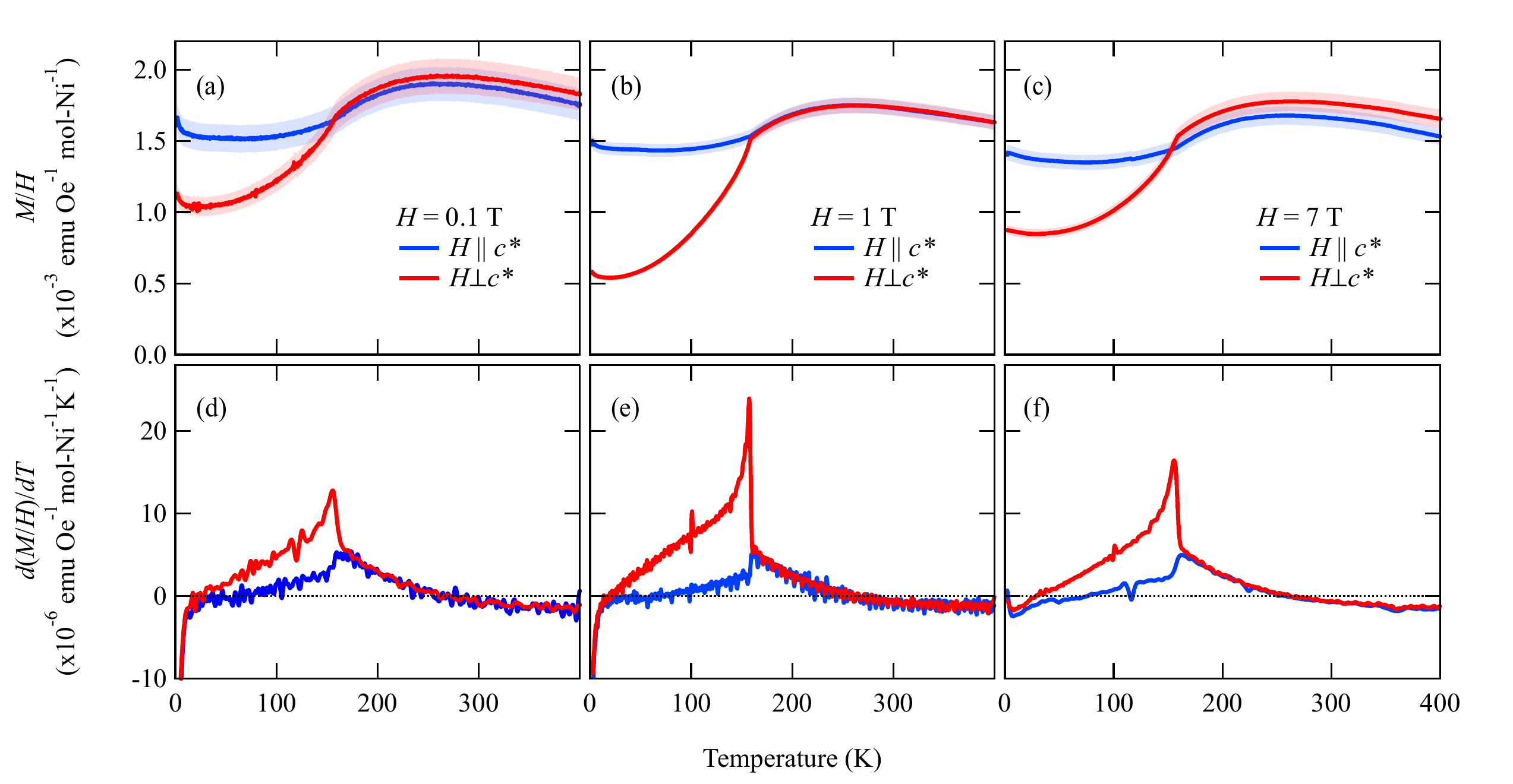}
    {\caption{\label{fig:fig_1_MoverH_vs_temp}Normalized magnetization as function of temperature $M/H(T)$ of a Ni$_2$P$_2$S$_6$ crystal for external fields of 0.1\,T (a), 1\,T (b), and 7\,T (c). Experimental uncertainty is show as a lightly colored band around each curve. (d-f) First derivatives of the respective $M/H(T)$ curves shown in (a-c).}}
\end{figure*}

The family of transition metal chalcogenophosphate $M$P$X_3$ materials---where $M = {}$Mn, Fe, Co, and Ni, to name a few (see~\cite{Grasso_2002_MPX3_family} for a comprehensive list), and $X = {}$S or Se---hosts a wide variety of fascinating physical properties. These materials crystallize in the monoclinic \textit{C2/m} spacegroup~\cite{Ouvrard_1985_MPS3_sx_XRD}, and are all semiconductors at ambient pressure, with band gaps larger than 1\,eV~\cite{Grasso_2002_MPX3_family}. They exhibit a particular structural motif; a P dimer sits symmetrically at the center of the transition metal hexagon, with each P covalently bonded to its three neighboring S atoms, forming a [P$_2$S$_6$]$^\mathrm{4-}$ cluster~\cite{Brec_1986_MPS3_review, Piacentini_1984_XAS_FeNiPS3}. Therefore, we will hereafter refer to these compounds using the doubled formula $M_2$P$_2X_6$. 

The $M_2$P$_2X_6$ materials were investigated both in the context of low dimensional materials physics and Li-based battery applications~\cite{Brec_1986_MPS3_review}. More recently, Fe$_2$P$_2$S$_6$ was shown to undergo a metal-insulator transition and two structural transitions at high pressure, consistent with expectations for a Mott or charge transfer insulator~\cite{Haines_2018_FePS3_pressure}. Furthermore, a variety of spectroscopic techniques, combined with density functional theory, suggest that Ni$_2$P$_2$S$_6$ is a negative charge transfer insulator~\cite{Kim_2018_NiPS3_correlations, Zaanen_1985_gaps_el_struc_TMs}. The magnetic properties of the $M_2$P$_2X_6$ family are strongly influenced by the transition metal element $M$. For example, substitution over the series including Mn, Fe, Co, and Ni results in a monotonic increase of the N{\'e}el temperature ($T_N$), with values of 82\,K, 116\,K, 122\,K, and 155\,K, respectively~\cite{Brec_1986_MPS3_review}. The magnetic structure is also modified with substitution, with the relevant case of Ni$_2$P$_2$S$_6$ found to display zig-zag antiferromagnetic order with the moment direction canted slightly out of the plane, mostly along the crystalline $a$ direction with wave vector $\mathbf{k}=\left[010\right]$.~\cite{Brec_1986_MPS3_review, Taylor_1973_MPS3_synthesis_props, Joy_1992_MPS3_magnetism, Wildes_2015_NiPS3_neutron_magstruc, Lancon_2018_NiPS3_INS}.

Several ${}^{31}$P  nuclear magnetic resonance (NMR) studies were conducted on Ni$_2$P$_2$S$_6$ and related systems, focused mostly on powder samples~\cite{Berthier_1980_intercalation_NMR, Berthier_1978_MPS3_NMR_Li, Ziolo_1988_31P_NMR_MPX3, Torre_1989_MPX3_31P_NMR}. The effects of Li intercalation were also studied via NMR, but indicated relatively low Li mobility considering the electrochemical activity~\cite{Berthier_1980_intercalation_NMR, Berthier_1978_MPS3_NMR_Li}. Some of these investigations indicated that ${}^{31}$P NMR is sensitive to spin fluctuations via the hyperfine field produced by the transition metal local moments. Furthermore, NMR spin--lattice relaxation rate ($T_1^{-1}$) measurements indicated strong field dependence of $T_N$~\cite{Ziolo_1988_31P_NMR_MPX3, Torre_1989_MPX3_31P_NMR}. The authors Torre et al.\ found that the NMR shift ${}^{31}K$ was proportional to the magnetic susceptibility $\chi$\cite{Torre_1989_MPX3_31P_NMR}.

Here we present significantly more precise ${}^{31}$P NMR data on well-characterized single crystals of Ni$_2$P$_2$S$_6$. We find evidence of NMR's sensitivity to quasi-2D magnetic correlations via deviation of the NMR shift ${}^{31}K$ from proportionality to the bulk magnetic susceptibility $\chi$. We rule out the possible confounding mechanism of temperature-dependent crystalline electric field energy level depopulation on the hyperfine coupling via quantum chemistry calculations. We also find appreciable in-plane transferred hyperfine coupling between the P nuclei and the Ni moments, which is consistent with both Ni--S--Ni super-exchange and Ni--S--S--Ni super-super-exchange as a likely explanation for the large nearest-neighbor and third-nearest-neighbor exchange couplings $J_1$ and $J_3$, respectively~\cite{Lancon_2018_NiPS3_INS}. In contrast to the literature, we observe no evidence to indicate field dependence of $T_N$ (up to 7\,T via magnetization, and 12\,T via NMR $T_1^{-1}$). In the magnetic state $T_1^{-1}$ follows a $T^5$ power law, indicating that a three-magnon process dominates the relaxation. Our spectral data in the antiferromagnetic state provide strong evidence for the existence of stacking faults, where the orientation of the layers is rotated in 60 degree increments.

\section{Experimental Methods}
\label{sec:exp_thry_details}

Single crystals of Ni$_2$P$_2$S$_6$ were grown by the chemical vapor transport technique, using iodine as a transport agent~\footnote{{S. Selter, Y. Shemerliuk, M.-I. Sturza, A. U. B. Wolter, B. B{\"u}chner, and S. Aswartham, Evolution of magnetic anisotropy in 2D van der Waals (Fe$_{1-x}$Ni$_x$)$_2$P$_2$S$_6$ single crystals, 2020 (unpublished)} and~\cite{Wildes_2015_NiPS3_neutron_magstruc}.} to obtain shiny plate-like crystals with dimensions of up to 2\,$\times$\,2\,$\times$\,0.2\,mm. The crystals were thoroughly characterized structurally by single crystal X-ray diffractometry (scXRD) and regarding the chemical composition by scanning electron microscopy (SEM) using a backscattered electron detector (BSE) and energy dispersive X-ray spectroscopy (EDX) (see Appendix~\ref{sec:elemental_comp_and_xrd} for further details).

The DC magnetization $M$ was measured as function of temperature $T$ and field $H$ using a superconducting quantum interference device vibrating sample magnetometer (SQUID-VSM) from Quantum Design. For comparison with NMR, the magnetic susceptibility was derived from magnetization data taken at 1\,T. $M$ vs $H$ was verified to be linear and independent of crystal orientation in the normal state to within the experimental error.

Two crystals of Ni$_2$P$_2$S$_6$ were selected for NMR measurements: hereafter referred to as crystal A and crystal B. The crystals display identical angular dependence of the normal state shift and relaxation rate, indicating homogeneity across samples. Both crystals were plate-like with well-defined facets and masses of approximately 1~mg. All shift values were calculated with respect to a ${}^{31}$P standard sample of 85\% H$_3$PO$_4$ in water.

NMR experiments were conducted in a 7\,T superconducting magnet with a homogeneity of better than 1\,ppm over a 1\,cm diameter spherical volume. A home-built probe with a single-axis goniometer was used for sample rotation and alignment. The sample temperature was controlled using a flow cryostat from Janis (sample in helium gas) with a calibrated Lakeshore Cernox temperature sensor. NMR measurements were performed with an Apollo spectrometer from Tecmag. A standard spin-echo pulse sequence ($\frac{\pi}{2}$--$\pi$) was used for spectral measurements, and an inversion-recovery pulse sequence ($\pi$--$\frac{\pi}{2}$--$\pi$) was used to measure $T_1^{-1}$.

%
%

\section{Magnetization}
\label{sec:magnetization}

\begin{figure}
    \includegraphics[trim=0cm 0cm 0cm 0cm, clip=true, width=\linewidth]{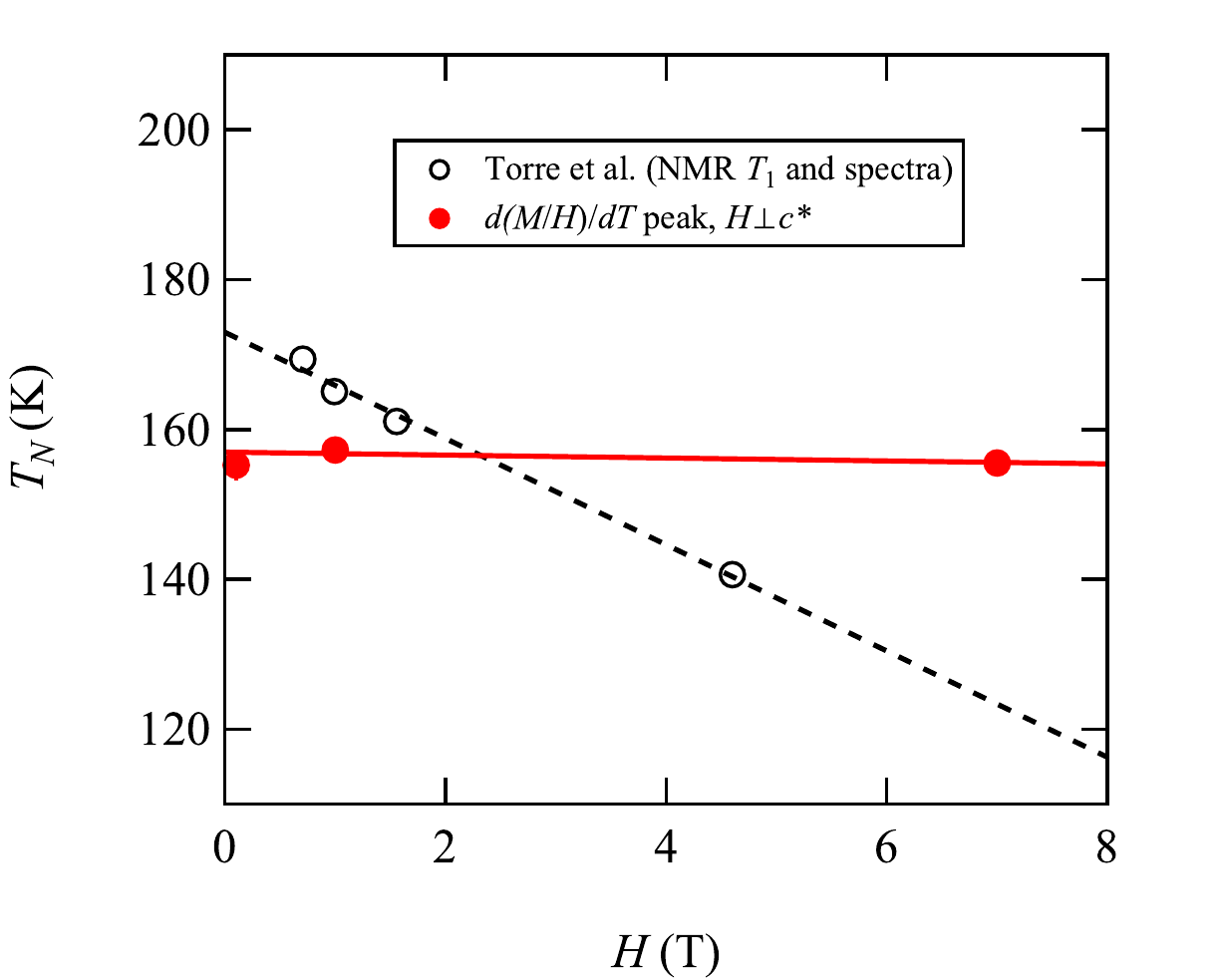}
    {\caption{\label{fig:fig_2_TN_vs_field}The N{\'e}el temperature $T_N$ as a function of field $H$ extracted from the peak in the derivative of $M/H(T)$ for $H\perp c^*$. Open circles were extracted from Torre et al.\ \cite{Torre_1989_MPX3_31P_NMR}.}}
\end{figure}

The magnetization divided by the applied field as a function of temperature $M/H(T)$ of a Ni$_2$P$_2$S$_6$ crystal is shown in Fig.~\ref{fig:fig_1_MoverH_vs_temp}. We extract the N{\'e}el temperature $T_N = 156 \pm 2$\,K from the sharp peak in the derivative of $M/H$ with respect to temperature for $H \perp c^*$, as shown in Fig.~\ref{fig:fig_1_MoverH_vs_temp}(d-f). In contrast to previous NMR measurements, $T_N$ is field independent over the range 0.1--7\,T, as shown in Fig.~\ref{fig:fig_2_TN_vs_field}. 

In three-dimensional (3D) antiferromagnets, $T_N$ is typically ascribed to the maximum value of $M/H(T)$. However, the \textit{inflection point} of $M/H(T)$ is actually a more precise measure of $T_N$. In 3D systems these two features occur at nearly the same temperature. In quasi-low-dimensional materials, like Ni$_2$P$_2$S$_6$, the reduced dimensionality of the interactions leads to a reduction of the long-range magnetic ordering temperature. A short range correlated regime emerges in between the ordered state and the uncorrelated paramagnetic state. As a result, the maximum in $M/H(T)$ is no longer a valid measure of the ordering temperature and the inflection point must be used to define $T_N$.

$M/H$ is isotropic in the paramagnetic state to within the experimental error, and deviations therefrom have been shown to be due to strain induced by gluing the samples onto the sample holder~\cite{Wildes_2015_NiPS3_neutron_magstruc}. A broad maximum, centered at $T_\mathrm{max} = 262 \pm 5$\,K, is attributed to the emergence of short-range spin correlations~\cite{de_jongh_2001_exp_mag_mod_sys}. The suppression  of $M/H$ with decreasing temperature below $T_\mathrm{max}$ indicates that antiferromagnetic interactions are dominant.

Below $T_N$, $M/H$ becomes anisotropic with respect to the external field direction. Furthermore, $M/H$ for $H \perp c^*$ is found to be significantly smaller than for $H \parallel c^*$. Accordingly, the antiferromagnetic easy axis is expected to lay in the $ab$-plane while the $c^*$-direction is a magnetic hard axis, in agreement with literature~\cite{Wildes_2015_NiPS3_neutron_magstruc, Joy_1992_MPS3_magnetism}.

\begin{figure}
    \includegraphics[trim=0cm 0cm 0cm 0cm, clip=true, width=\linewidth]{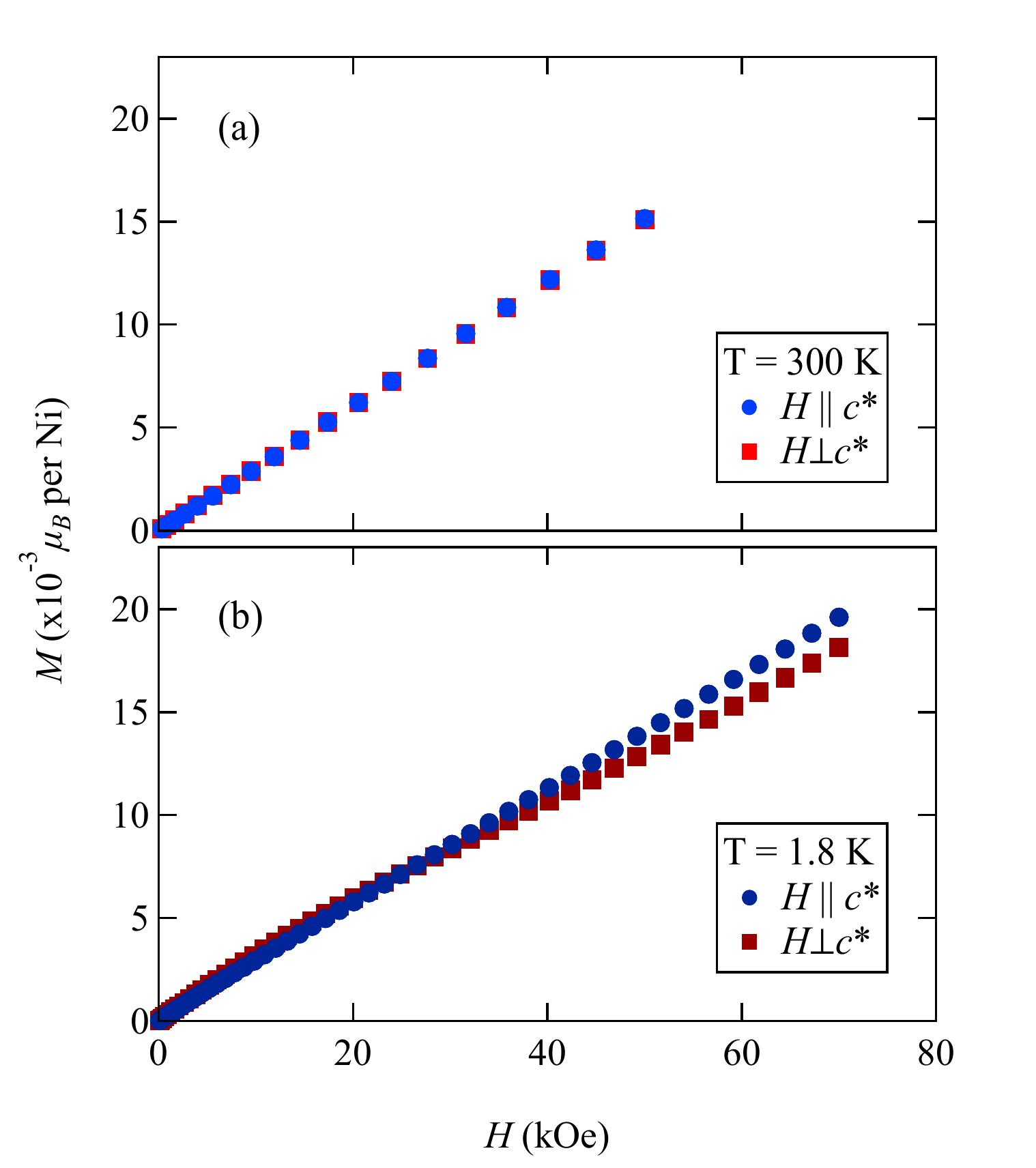}
    {\caption{\label{fig:fig_3_M_vs_H}Magnetization as function of external field of a Ni$_2$P$_2$S$_6$ crystal at 300\,K (a) and 1.8\,K (b).}}
\end{figure}

Comparing the temperature dependencies for the different fields shows no significant influence of external fields in the paramagnetic and short range correlated regime. This is corroborated by the linearity of $M$ vs $H$ at 300\,K, shown in Fig.~\ref{fig:fig_3_M_vs_H}(a). In the magnetically ordered state a small deviation between 0.1\,T and 7\,T in-plane is observed, while the measurements out-of-plane match. Measurements at 1\,T, with careful orientation of the crystalline axes with respect to the applied field, show that the deviation between 0.1\,T and 7\,T is due to a slight misalignment. At the lowest temperatures, a Curie-like tail is found for 0.1\,T, which is suppressed at 7\,T. Such a tail is attributed to weak ferromagnetic contributions which may be caused by crystal defects.

This weak contribution is also observed at small fields up to approximately 1\,T as a change of the slope in the field dependence of $M$ measured at 1.8\,K, as shown in Fig.\,\ref{fig:fig_3_M_vs_H}(b). However at higher fields, a linear dependence between external field and magnetization is found up to the highest measured field 7\,T. Therefore, if a spin-flop transition exists in Ni$_2$P$_2$S$_6$, the field at which it occurs is larger than 7\,T.

\section{Quantum chemistry calculations}
\label{sec:quantum_chemistry}

\begin{table}
\begin{ruledtabular}
\begin{tabular}{lcc}
Ni 3 \textit{d}$^{8}$ state                                   & CASSCF (eV)      & MRCI (eV)        \\
\hline \\[-1.5ex]
${}^3$\textit{A}$_{2g}$ (\textit{t}$_{2g}^6$\textit{e}$_g^2$) & 0.00             & 0.00             \\[1ex]
${}^3$\textit{T}$_{2g}$ (\textit{t}$_{2g}^5$\textit{e}$_g^3$) & 0.84, 0.84, 0.89 & 0.99, 1.00, 1.05 \\[1ex]
${}^3$\textit{T}$_{1g}$ (\textit{t}$_{2g}^5$\textit{e}$_g^3$) & 1.47, 1.48, 1.49 & 1.71, 1.73, 1.74 \\[1ex]
${}^1$\textit{E}$_g$ (\textit{t}$_{2g}^6$\textit{e}$_g^2$)    & 2.17, 2.18       & 2.05, 2.05       \\[1ex]
${}^1$\textit{T}$_{2g}$ (\textit{t}$_{2g}^5$\textit{e}$_g^3$) & 2.95, 2.96, 3.05 & 3.01, 3.02, 3.10 \\[1ex]
${}^3$\textit{T}$_{1g}$ (\textit{t}$_{2g}^4$\textit{e}$_g^4$) & 3.34, 3.36, 3.52 & 3.32, 3.33, 3.52 \\[1ex]
${}^1$\textit{A}$_{1g}$ (\textit{t}$_{2g}^6$\textit{e}$_g^2$) & 3.47             & 3.34             \\[1ex]
${}^1$\textit{T}$_{1g}$ (\textit{t}$_{2g}^5$\textit{e}$_g^3$) & 3.79, 3.81, 3.86 & 3.79, 3.81, 3.87 \\[1ex]
${}^1$\textit{T}$_{2g}$ (\textit{t}$_{2g}^4$\textit{e}$_g^4$) & 4.39, 4.40, 4.45 & 4.56, 4.57, 4.61 \\[1ex]
${}^1$\textit{E}$_g$ (\textit{t}$_{2g}^4$\textit{e}$_g^4$)    & 4.58, 4.58       & 4.76, 4.76       \\[1ex]
${}^1$\textit{A}$_{1g}$ (\textit{t}$_{2g}^6$\textit{e}$_g^2$) & 8.58             & 8.05             \\[0.5ex]
\end{tabular}
\end{ruledtabular}
{\caption{\label{tab:qc_multiplet_structure}Ni 3\textit{d}${}^{8}$ multiplet structure as computed by \textit{ab initio} quantum chemistry for NiPS$_{3}$.}}
\end{table}

Nickel commonly comes with a 2+ ionization state in chalcogenides. The lower-lying features at 1.1 and 1.7 eV in the optical absorption spectrum of Ni$_2$P$_2$S$_6$~\cite{Kim_2018_NiPS3_correlations} are in rather good correspondence with the low-energy \textit{d}-\textit{d} transitions in the Ni$^{2+}$ prototype material La$_2$NiO$_4$~\cite{Fabbris_2017_doping}, suggesting indeed a 2+ valence state. The magnetic properties of Ni$_2$P$_2$S$_6$ were also interpreted in terms of \textit{S}=1 Ni$^{2+}$ ions~\cite{Chandrasekharan_1994_magnetism, Wildes_2015_NiPS3_neutron_magstruc, Lancon_2018_NiPS3_INS, Kim_2019_suppression}, with a zero-field splitting on the order of 1~meV for the Ni$^{2+}$ \textit{t}$_{2g}^6$\textit{e}$_g^2$ ground-state configuration~\cite{Chandrasekharan_1994_magnetism, Lancon_2018_NiPS3_INS,Kim_2019_suppression}. In this context, we performed quantum chemical electronic-structure calculations, on an atomic fragment consisting of one reference NiS$_6$ octahedron along with three nearest-neighbor octahedra sharing edges with the reference unit and three adjacent P$_2$ dimers~\footnote{All-electron triple-$\zeta$ basis sets (BS's) with polarization functions were used for the central NiS$_{6}$ octahedron, of Douglas-Kroll-type for Ni. The adjacent Ni$^{2+}$ ions were modeled as closed-shell Zn$^{2+}$ total-ion potentials provided with two $s$ functions while for the remaining ligands coordinating these cations we employed effective core potentials (ECP's) and valence BS's of double-$\zeta$ quality. ECP's and valence BS's of double-$\zeta$ quality were also used for the closest P species around the central NiS$_{6}$ unit. The quantum chemical package {\sc molpro} \cite{Werner_2012_Molpro} was employed. All ECP's and BS's were taken from the {\sc molpro} library.}. The remaining part of the extended crystalline surroundings was modeled as an effective electrostatic field.

The total number of electrons assigned to this atomic fragment was chosen according to the commonly accepted picture of Ni${}^{2+}$ ions and [P$_2$S$_6$]$^{4-}$ entities in Ni$_2$P$_2$S$_6$. Through quantum chemical complete-active-space self-consistent field (CASSCF) calculations~\cite{Helgaker_2014_molecular}, we confirm the peculiar P-P chemical bond with a doubly occupied 3\textit{s}-3\textit{s} bonding orbital. Using for simplicity notations corresponding to cubic octahedral symmetry, a ${}^{3}$\textit{A}$_{2g}$ \textit{t}$_{2g}^6$\textit{e}$_g^2$ ground state is found for the central Ni site. The actual point group symmetry is however lower since the ligand cage around a given Ni ion features some amount of trigonal compression and additional small distortions that yield three sets of slightly different Ni-S bond lengths. This is the reason second-order spin-orbit interactions give rise to zero-field splitting and single-ion anisotropy.

Results of both CASSCF and multireference configuration-interaction (MRCI)~\cite{Helgaker_2014_molecular} calculations are listed for the Ni$^{2+}$ \textit{d}$^8$ multiplet structure in Table~\ref{tab:qc_multiplet_structure}. The CASSCF optimization was carried out for an average of all triplet and singlet states arising from the $d^8$ configuration. The MRCI treatment implies single and double excitations out of the central-octahedron S 3\textit{p} and Ni 3\textit{d} orbitals on top of the CASSCF expansion and brings corrections of up to 0.25 eV to the relative energies. The $^3$\textit{A}$_{2g}$-$^3$\textit{T}$_{2g}$ and $^3$\textit{A}$_{2g}$-$^3$\textit{T}$_{1g}$ splittings, for example, are significantly enlarged since the leading ground-state configuration \textit{t}$_{2g}^6$\textit{e}$_g^2$ entails less charge within the 3\textit{d} $\sigma$-like \textit{e}$_g$ levels and the ${}^3$\textit{A}$_{2g}$ wavefunction undergoes therefore stronger renormalization when S 3\textit{p} to Ni 3\textit{d} charge-transfer effects are accounted for by MRCI (see also discussion in Refs.~\cite{Hozoi_2009_spin,Hozoi_2011_ab_initio}). The lowest MRCI excitation energies, ${}^3$\textit{A}$_{2g}$-${}^3$\textit{T}${}_{2g}$ and ${}^3$\textit{A}$_{2g}$-${}^3$\textit{T}$_{1g}$, are in fact in good agreement with transitions at 1.1 and 1.7 eV in optical absorption~\cite{Kim_2018_NiPS3_correlations}. When spin-orbit couplings are accounted for as well in MRCI, according to the procedure described in Ref.~\cite{Berning_2000_spin}, a zero-field splitting of 0.7~meV is computed (not shown in the table), with easy-plane anisotropy. More details in this regard will be provided elsewhere.

\section{NMR Results}
\subsection{Normal state spectral measurements and shift anomaly}
\label{subsec:normal_state_spectra_Kchi_anomaly}

\begin{figure} 
    \includegraphics[trim=0.25cm 0cm 0.25cm 0cm, clip=true, width=\linewidth]{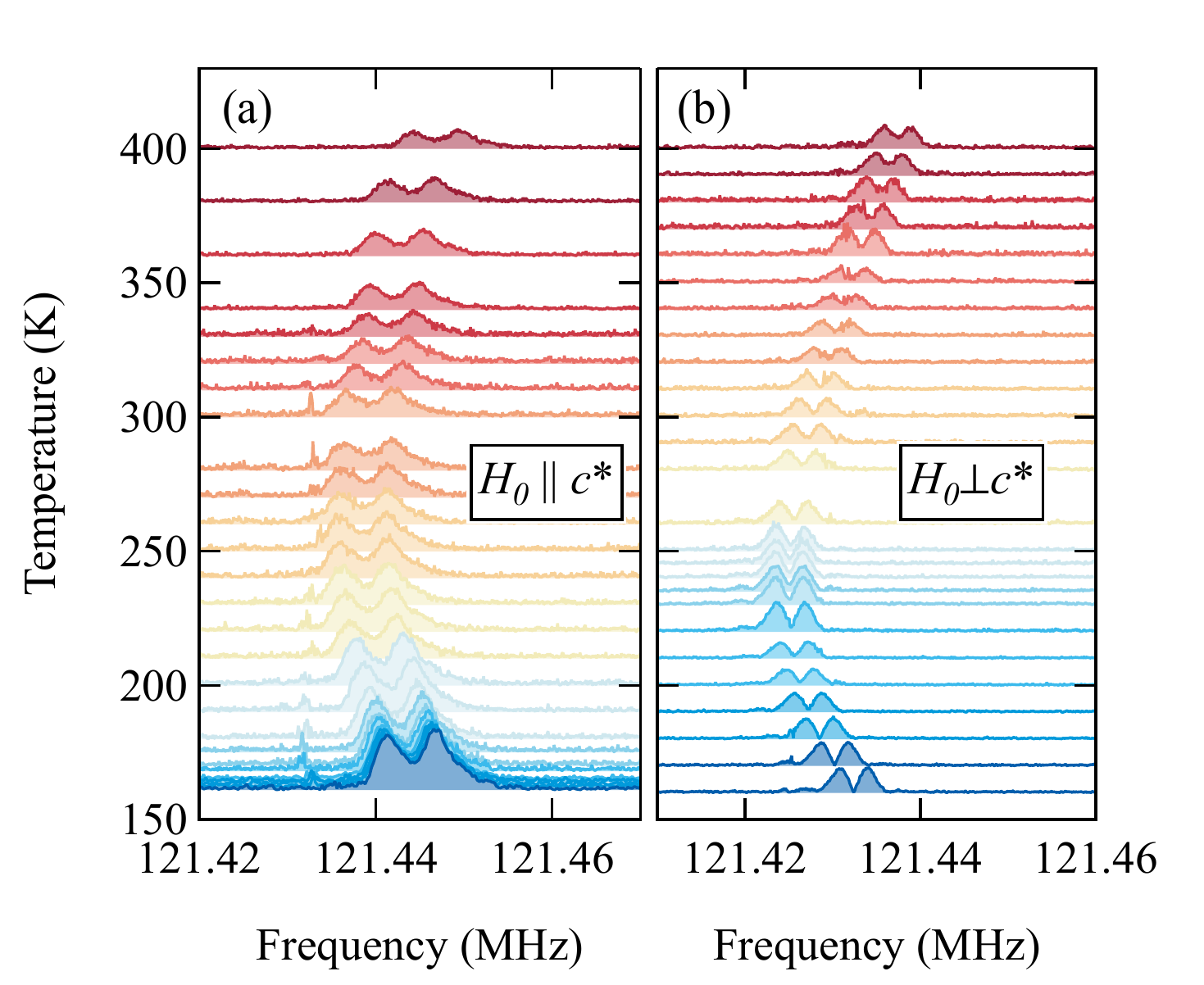}
    {\caption{\label{fig:fig_4_spectra_vs_temp}${}^{31}$P NMR spectra from crystal B for 160\,K $< T <$ 400\,K (offset by temperature) with $H_0 \parallel c^*$ (a) and $H_0 \perp c^*$ (b).}}
\end{figure}

\begin{figure} 
    \includegraphics[trim=0.5cm 0cm 0.5cm 0cm, clip=true, width=\linewidth]{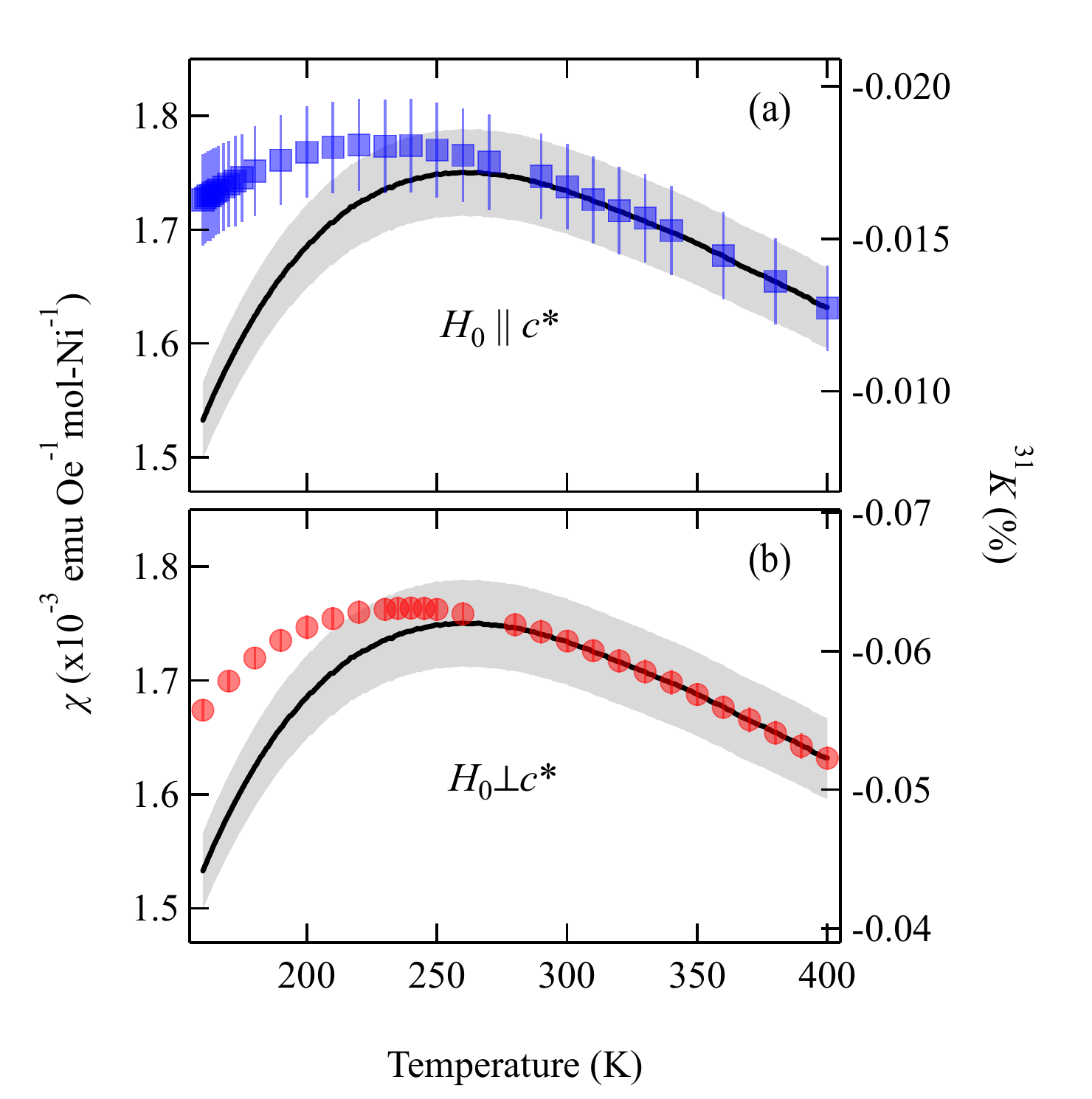}
    {\caption{\label{fig:fig_5_K_and_chi_vs_temp}Magnetic susceptibility $\chi$ (left axis, solid lines) and NMR shift ${}^{31}K$ (right axis, markers) as a function of temperature for $H_0 \parallel c^*$ (a) and $H_0 \perp c^*$ (b). The right axis is scaled to illustrate the breakdown in proportionality below approximately 275\,K. The grey shaded band indicates the uncertainty of $\chi$.}}
\end{figure}

Our measurements of the temperature dependencies of the normal-state ${}^{31}$P NMR spectra are shown in Fig.~\ref{fig:fig_4_spectra_vs_temp}. There are two notable features about the spectral measurements: first, although there is only one P site, the spectra are double peaked for both orientations of the crystal with respect to the magnetic field. Second, the temperature dependencies of the spectra yield a nonmonotonic NMR shift $K$.

\begin{figure*} 
    \includegraphics[trim=0cm 0cm 0cm 0cm, clip=true, width=\linewidth]{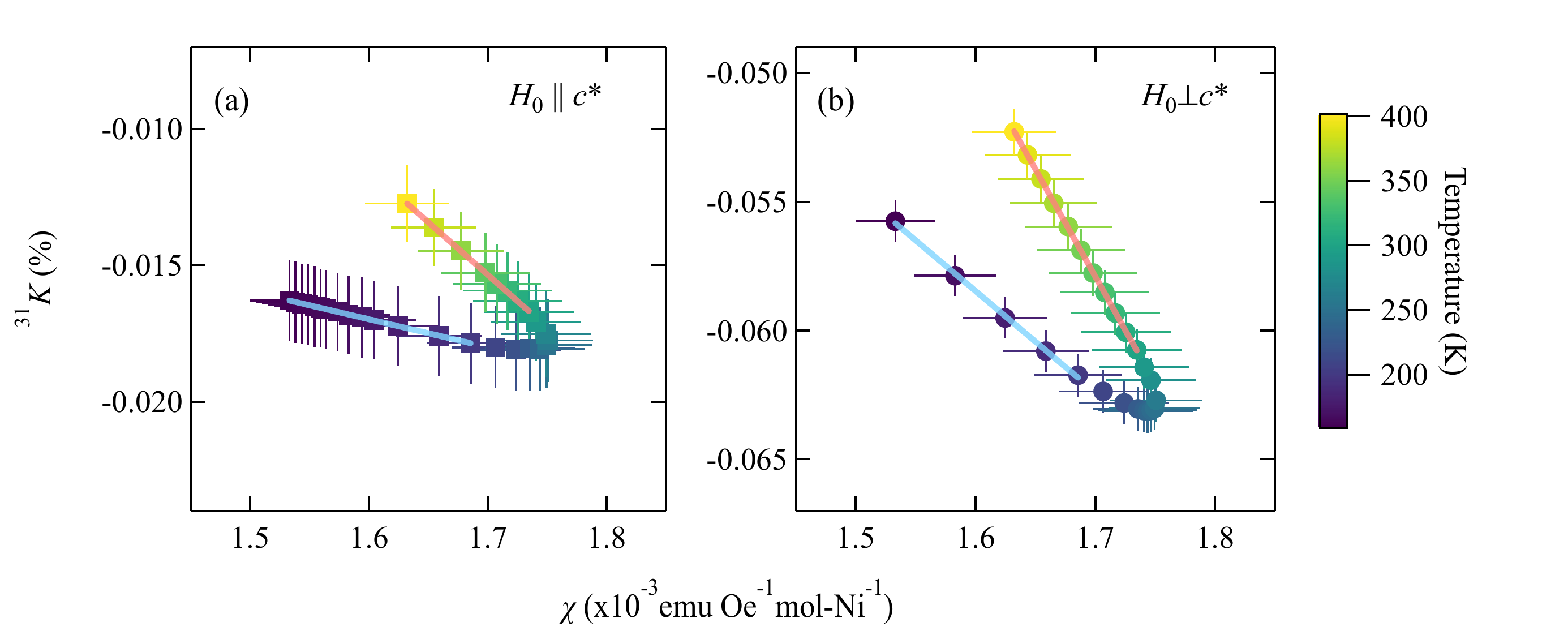}
    {\caption{\label{fig:fig_6_K_vs_chi_layout}${}^{31}K$ vs $\chi$ for $H_0 \parallel c^*$ (a) and $H_0 \perp c^*$ (b) with temperature as an implicit parameter encoded via the color scale on the right. Solid lines are linear fits to extract the hyperfine coupling constants, which are detailed in Table~\ref{tab:hf_coupling_exp}.}}
\end{figure*}

With respect to the former, the spectral splitting agrees well with the expected spectral profile for a ``Pake doublet,'' a phenomenon resulting from the previously mentioned P dimer~\cite{Pake_1948_pake_doublet}. The splitting is a result of the nuclear dipole-dipole interaction due to the close proximity of the P in the dimer. We draw this conclusion based on the expected angular dependence of the spectral splitting as shown in Appendix~\ref{sec:pake_doublet}. Considering the excellent agreement with theory, we subtract the effect of this interaction by considering the center of gravity of the spectrum for the rest of the article, as shown in the extracted NMR shift vs temperature in Fig.~\ref{fig:fig_5_K_and_chi_vs_temp}. At this point we note that, due to the very small value and temperature dependence of the shift, it was necessary to correct for the effects of macroscopic magnetism/shape anisotropy (see Appendix~\ref{sec:mac_mag_correct} for further details).

The second notable feature---the nonmonotonic temperature dependence of ${}^{31}K$---leads us to, what is arguably, the most exciting result of this study (summarized in Fig.~\ref{fig:fig_5_K_and_chi_vs_temp}. The figure shows the magnetic susceptibility $\chi$ on the left axis and ${}^{31}K$ for on the right axis a function of temperature for both $H_0 \perp c^*$ and $H_0 \parallel c^*$. In an uncorrelated paramagnetic system, the bulk magnetic susceptibility must scale with the NMR shift. However, we find that below approximately 275\,K the scaling breaks down, with $\chi$ decreasing more strongly than the NMR shift increases.

This anomaly can be investigated in further detail by plotting the NMR shift as a function of the magnetic susceptibility with temperature as an implicit parameter in a so-called Clogston-Jaccarino plot~\cite{Clogston_1964_CJ_KvsChi}. This analysis is shown for the crystal oriented with both $H_0 \parallel c^*$ and $H_0 \perp c^*$ in Fig.~\ref{fig:fig_6_K_vs_chi_layout}. Solid lines are fits to extract the hyperfine couplings for both the high and low temperature linear regimes, which are tabulated in Table~\ref{tab:hf_coupling_exp}. The small magnetic moment of the crystal resulted in nontrivial uncertainty in the $\chi$ data. Therefore, ${}^{31}K$ vs $\chi$ fitting was performed using an orthogonal distance regression algorithm in Igor Pro to take into account the uncertainty of both the $y$ (${}^{31}K$) and $x$ ($\chi$) data sets.

\begin{figure} 
    \includegraphics[trim=0cm 0cm 0cm 0cm, clip=true, width=\linewidth]{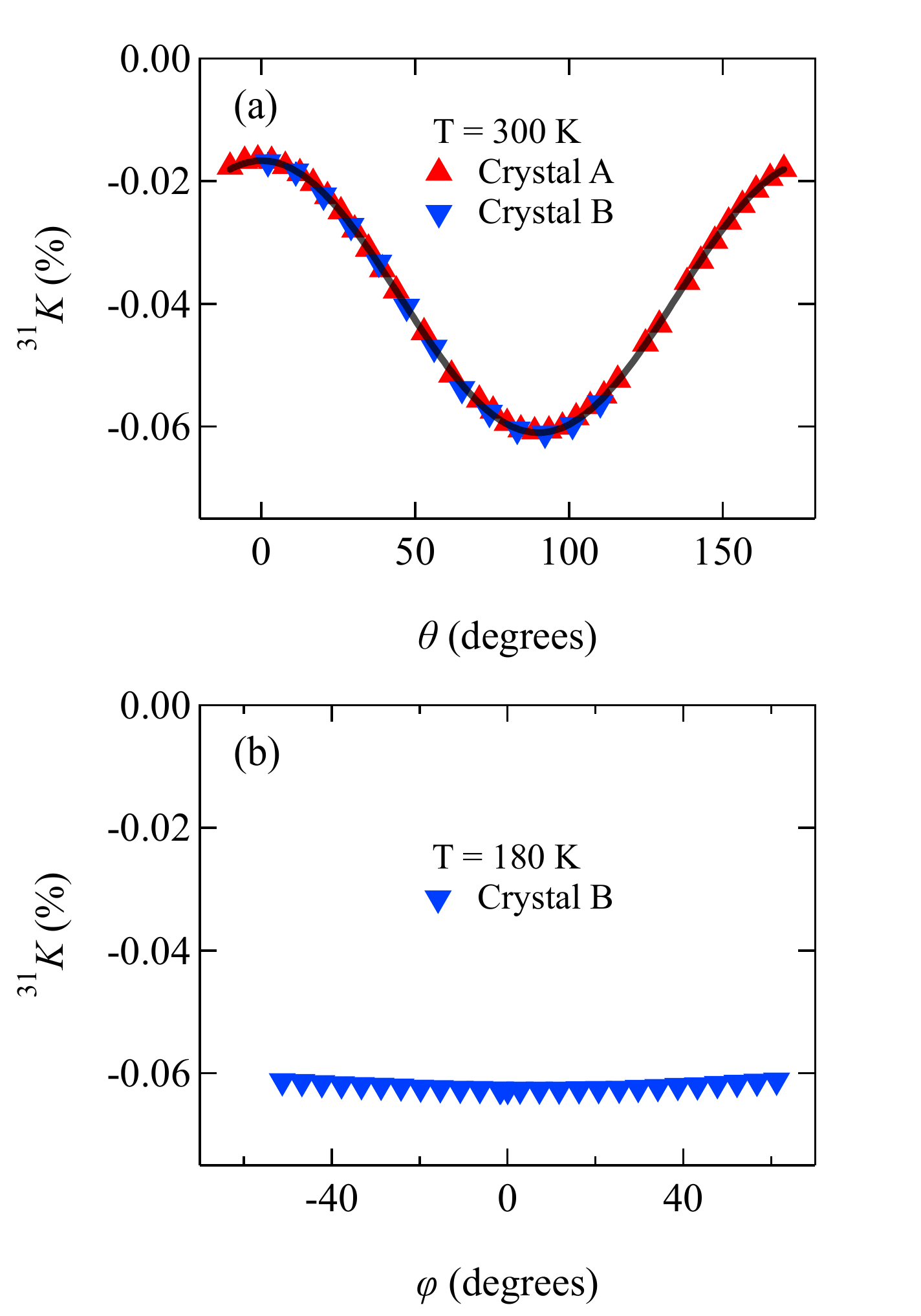}
    {\caption{\label{fig:fig_7_K_normal_ang_dep}(a) Out-of-plane angular dependence of the NMR shift at $T = 300$\,K of crystals A (upward-facing red triangles) and B (downward-facing blue triangles). The grey curve is a global fit to extract $K_{c^*}$ and $K_{ab}$ as described in the text. (b) Angular dependence of the NMR shift in the $a$--$b$ plane at $T = 180$\,K of crystal B.}}
\end{figure}

We measured the angular dependence of the ${}^{31}$P spectrum at $T = 300$\,K for out-of-plane rotation ($\theta$--transverse to the P--P dimer axis) and at $T = 180$\,K in-plane rotation ($\phi$--about the P--P dimer axis), shown in Fig.\ref{fig:fig_7_K_normal_ang_dep}. Once again, we corrected all angular-dependent data for effects arising from macroscopic magnetism (Appendix~\ref{sec:mac_mag_correct}) and for the spectral splitting of the Pake doublet (Appendix~\ref{sec:pake_doublet}). Taken together, these rotation data show that the ${}^{31}$P resonances arise from sites with axial symmetry, such that $K_a = K_b \neq K_{c^*}$. Hereafter, we will refer to the in-plane shift tensor components as $K_{ab} \equiv K_a = K_b$. The out-of-plane rotation experiment was conducted on both crystal A and crystal B. We fit both data sets globally to the equation for the angular dependence of the shift
\begin{equation}
K(\theta) = K_\mathrm{iso} + K_\mathrm{ax}\left(3\cos^2{\left((\theta - \theta_0)\frac{\pi}{180}\right)} - 1\right),
\end{equation}
where $K_\mathrm{iso} = \frac{1}{3}(2K_{ab} + K_{c^*})$ and $K_\mathrm{ax} =  \frac{1}{3}(K_{c^*} - K_{ab})$. The resulting fit is shown as a dark grey curve in Fig.~\ref{fig:fig_7_K_normal_ang_dep}(a). The extracted shift tensor elements are as follows: $K_{c^*} = -0.0166$\,\%, $K_{ab} = -0.0610$\,\%. The uncertainty of the shift values is $\pm~0.0002$\,\%.

\subsection{Spin--lattice relaxation rate}
\label{subsec:spin--lattice_relaxation}

\begin{figure}
    \includegraphics[trim=0cm 0cm 0.5cm 0cm, clip=true, width=\linewidth]{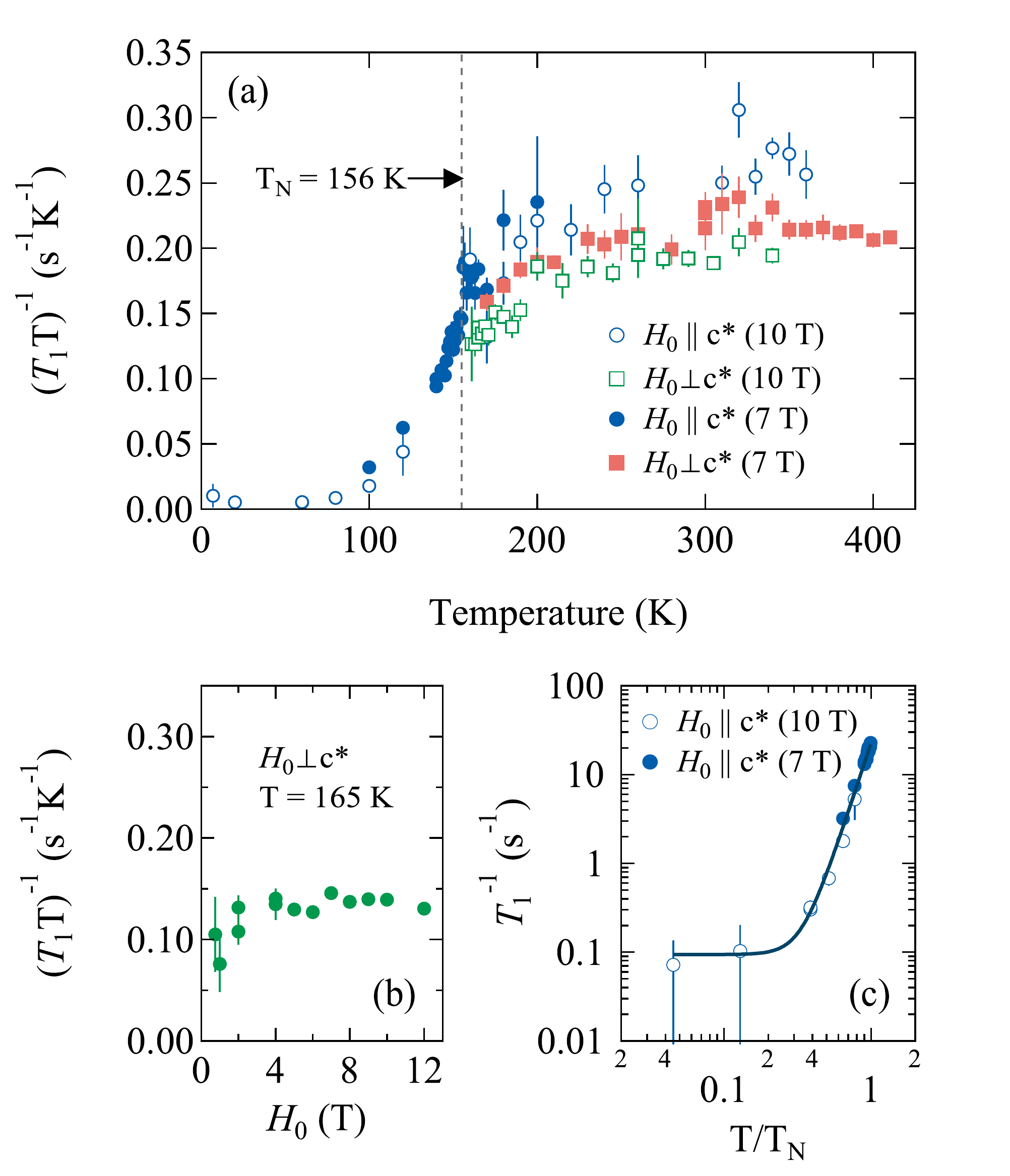}
    {\caption{\label{fig:fig_8_T1}(a) ${}^{31}$P Spin--lattice relaxation rate divided by temperature $(T_1T)^{-1}$ vs temperature for $H_0 \parallel c^*$ and $H_0 \perp c^*$ and $H_0 = 7$ and 10\,T. The N{\'e}el temperature $T_N$ is marked with a dashed vertical line. (b) $(T_1T)^{-1}$ vs applied magnetic field for $H_0 \perp c^*$ at $T = 165$\,K. (c) Spin--lattice relaxation rate $T_1^{-1}$ vs reduced temperature $T/T_N$.}}
\end{figure}

We measured $T_1^{-1}$ as a function of temperature for $H_0 \parallel c^*$ and  $H_0 \perp c^*$ with $H_0 = 7$ and 10\,T. The inversion recovery curves---i.e. the integrated phase-corrected real part of the spin echo vs the time between the inverting pulse and the spin-echo sampling pulses---were well fit by a single exponential relaxation function given by,
\begin{equation}
M(t) = M_0\left(1 - 2Fe^{-t/T_1}\right).
\end{equation}
In the above expression, $M_0$ is the equilibrium nuclear magnetization, $F$ is the inversion fraction, $t$ is the time between the inverting $\pi$ pulse and the spin-echo $\frac{\pi}{2}$--$\pi$ pulses, and $T_1$ is the spin--lattice relaxation time. These results are summarized in Fig.~\ref{fig:fig_8_T1}(a), plotted as $(T_1T)^{-1}$ vs temperature.

We also measured the field dependence of $(T_1T)^{-1}$ just above $T_N$ at $T = 165$\,K for $H_0 \perp c^*$. $(T_1T)^{-1}$ exhibits little to no field dependence as a function of applied field $H_0$, though it is possible that there is some suppression at the lowest fields. The standard errors of the fit parameters at the lowest fields become quite large. 

The behavior of $T_1^{-1}$ in the antiferromagnetic state is consistent with relaxation dominated by magnon scattering. A power law with a constant background term of the form,
\begin{equation}
T_1^{-1} = (T_1^{-1})_0 + bT^\alpha
\end{equation}
describes the temperature dependence of the $T_1^{-1}$ data well and is shown as a solid line in Fig.~\ref{fig:fig_8_T1}(c). We performed the displayed fit on both the 7 and 10\,T data sets together and find $\alpha = 5.0 \pm 0.1$.

\begin{table}
\begin{ruledtabular}
\begin{tabular}{ccc}
              & High Temperature (T/$\mu_B$)     & Low Temperature (T/$\mu_B$) \\[0.25ex]
\hline \\[-1.5ex]
$A_{c^*}$     & $-0.2 \pm 0.1$             & $-0.06 \pm 0.05$      \\[0.5ex]
$A_{ab}$      & $-0.5 \pm 0.2$             & $-0.22 \pm 0.07$      \\[0.5ex]
\end{tabular}
\end{ruledtabular}
\caption{\label{tab:hf_coupling_exp}Hyperfine coupling constants extracted from fits to ${}^{31}$K vs $\chi$ as shown in Fig~\ref{fig:fig_6_K_vs_chi_layout}.}
\end{table}

\subsection{Magnetic state spectral measurements}
\label{subsec:magnetic_state}

\begin{figure}
    \includegraphics[trim=0cm 0cm 1cm 0cm, clip=true, width=\linewidth]{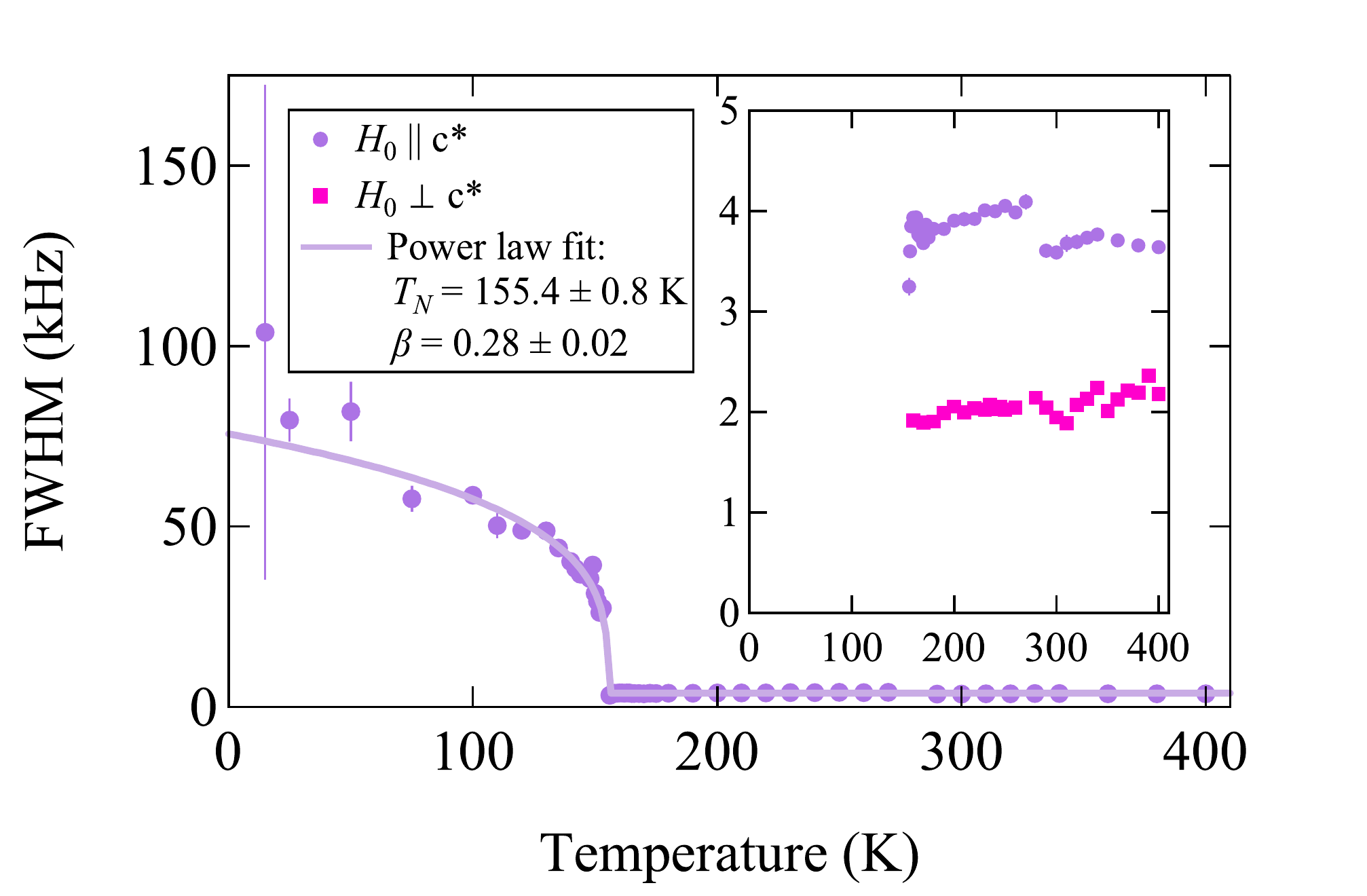}
    {\caption{\label{fig:fig_9_FWHM_vs_temp}Full width at half maximum (FWHM) of the ${}^{31}$P resonances vs temperature. The inset shows a zoomed-in view of the normal state FWHM.}}
\end{figure}

We extracted the temperature dependence of the linewidth of the spectra for $H_0 \parallel c^*$ and show the full width at half maximum (FWHM) in Fig.~\ref{fig:fig_9_FWHM_vs_temp}. The FWHM displays order-parameter-like behavior that we fit using a power law of the form,
\begin{equation}
\mathrm{FWHM}(T) = A(T_N - T)^\beta + \mathrm{FHMW}_0,
\end{equation}
where $A$ is a scaling parameter, $T_N$ is the N{\'e}el temperature, $\beta$ is the power law exponent, and $\mathrm{FHMW}_0$ is the normal state constant value of the FWHM. Our fit yields $T_N = 155.4 \pm 0.8$\,K---in close agreement with $T_N = 156$\,K extracted from our $M/H(T)$ measurements---and $\beta = 0.28 \pm 0.02$.

\begin{figure*}
    \includegraphics[trim=0cm 13.5cm 0cm 1cm, clip=true, width=\linewidth]{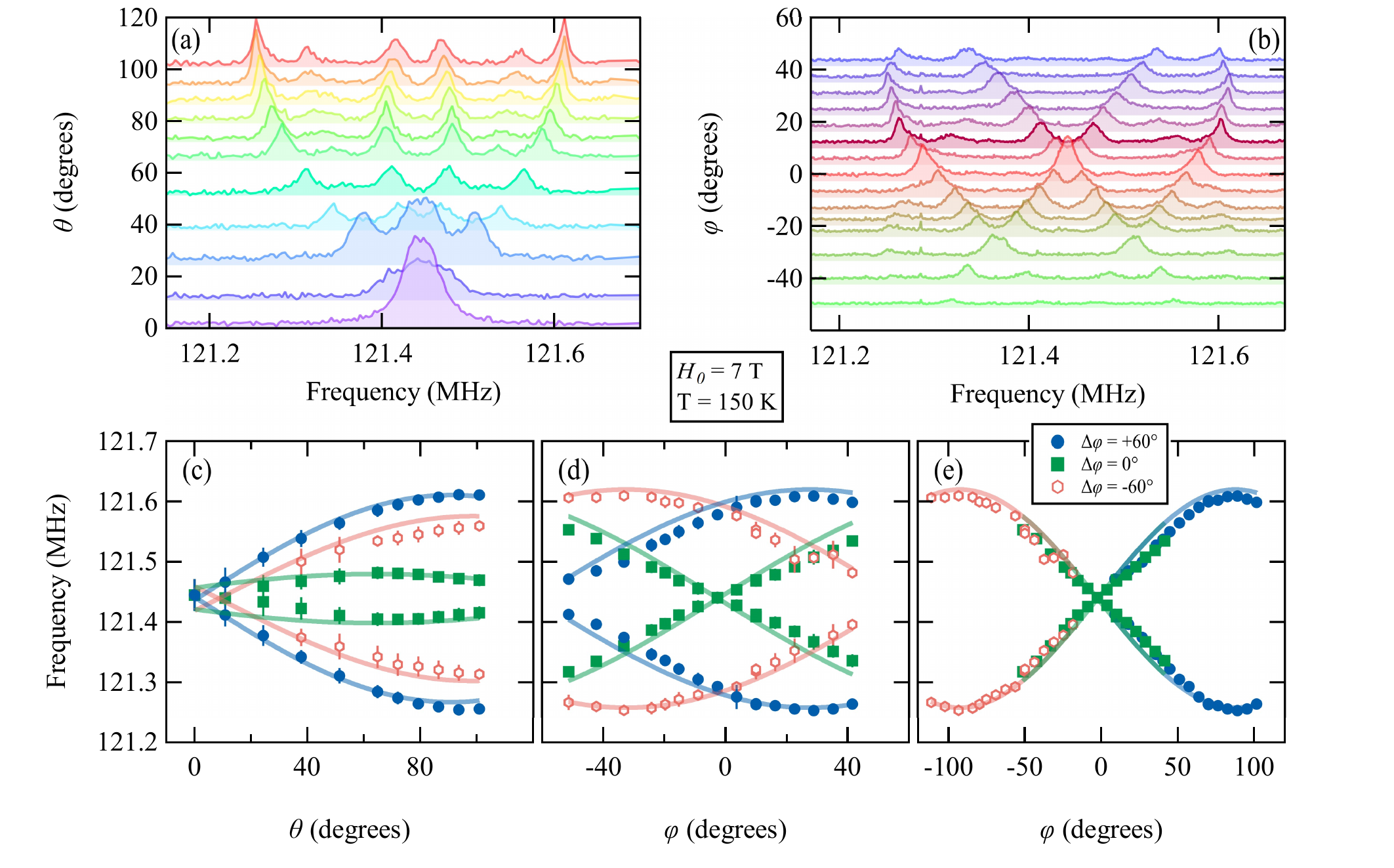}
    {\caption{\label{fig:fig_10_magnetic_state_ang_dep}${}^{31}$P NMR spectra offset by in-plane angle $\theta$ (a) and out-of-plane angle $\phi$ in the magnetic state at T=150\,K. (c) Frequencies vs $\theta$ extracted from multipeak fits to the spectra in (a). (d) Frequencies vs $\phi$ extracted from (b) by the same method. (e) In-plane rotation center frequencies from (d) offset by -60, 0, and 60 degrees. Blue, red, and green lines in (c-e) are simulations as discussed in the text.}}
\end{figure*}

We also measured the angular dependence of the ${}^{31}$P NMR spectrum in the magnetic state for both in-plane and out-of-plane rotation at $T = 150$\,K. These spectra and the extracted peak positions are shown in Fig.~\ref{fig:fig_10_magnetic_state_ang_dep}(a-d). The spectra contain three pairs of magnetically split peaks. The spectral weight of the three pairs of resonances, as designated by red open circles, blue closed circles, and green squares in Fig.~\ref{fig:fig_10_magnetic_state_ang_dep}(c-e) are as follows: $S_\mathrm{red} = 16$\,\%, $S_\mathrm{blue} = 37$\,\%, and $S_\mathrm{green} = 47$\,\%, with an uncertainty of $\pm 1$\,\%. The narrow resonance lines of the normal state spectrum disappear completely below $T_N$, indicating that no significant fraction of nuclei sample a magnetically disordered environment.

The angular dependence of the magnetic state spectra was used to extract the magnitude and orientation of the internal fields present at the ${}^{31}$P site. To take into account slight crystal misalignment with respect to the rotation and external magnetic field axes, the spectra were simulated via numerical exact diagonalization of the nuclear spin Hamiltonian including an internal hyperfine field interaction. The results of the simulations are shown as blue, green, and red lines in Fig.~\ref{fig:fig_10_magnetic_state_ang_dep}(c-e). Both in-plane and out-of-plane rotation experiments agree well with three sets of hyperfine field pairs that are offset with respect to each other by $-60$, 0, and 60 degrees.

\section{Discussion and Conclusions}
\label{sec:discussion_conclusion}

The two most perplexing points that merit discussion are the $K$--$\chi$ anomaly and the angular dependence of magnetic-state spectra. First, we will treat the $K$-$\chi$ anomaly and compare Ni$_2$P$_2$S$_6$ to similar systems where this effect has been observed. $A_2M$F$_4$, where $A = \mathrm{K~or~Rb}$ and $M = \mathrm{Mn,~Ni,~or~Co}$, are well known examples of quasi-2D antiferromagnets where the maximum in $\chi$ occurs well above the ordering temperatures~\cite{Breed_1967_2d_Mn_Fluorides, Breed_1969_Co_Fluorides, Maarschall_1969_K2NiF4_NMR, van_der_Klink_2010_La2NiO4p17_K2NiF4_NMR}. $T_N$ in quasi-2D antiferromagnets is observed to be significantly lower, in general, than the broad maximum in the susceptibility~\cite{Chung_2004_FeTa2O6_2D_AFM, de_jongh_2001_exp_mag_mod_sys}. This has been argued to be a result of short-range order that emerges above $T_N$ due to the reduced dimensionality~\cite{de_jongh_2001_exp_mag_mod_sys}.

To our knowledge, the $K$--$\chi$ anomaly in a quasi-2D magnetic system has been directly investigated in only three previous studies. In the first, van der Klink and Brom posit that in K$_2$NiF$_4$ ${}^{61}$Ni $K$ and $\chi$ are differently sensitive to the onset of short-range correlations above $T_N$~\cite{van_der_Klink_2010_La2NiO4p17_K2NiF4_NMR}. The second example is the case of VOMoO$_4$, in which the shift anomaly was concluded to stem from a frustration-induced structural transition, which resulted in significant changes to the hyperfine couplings~\cite{Carretta_2002_VOMoO4}. A third, very recent, example of a $K$--$\chi$ anomaly in a quasi-2D magnetic insulator is the case of the honeycomb lattice material Na$_2$IrO$_3$~\cite{Sarkar_2019_Na2IrO3_disorder}. Here Sarkar et al.\ discuss the role of disorder and/or interlayer correlations perturbed by stacking faults as a possible mechanism behind the $K$--$\chi$ anomaly in Na$_2$IrO$_3$~\cite{Sarkar_2019_Na2IrO3_disorder}. In all of the above mentioned cases, the shift anomaly was associated with line broadening, and therefore likely results from static short-range magnetic order. Neutron scattering measurements~\cite{Wiedenmann_1981_MnFePSe3_neutrons} also find evidence for short-range static magnetic order that persists at temperatures well above $T_N$ in Mn$_2$P$_2$S$_6$ and Fe$_2$P$_2$S$_6$.

The origin of the $K$--$\chi$ anomaly in Ni$_2$P$_2$S$_6$ is likely related to the onset of quasi-2D-induced short-range correlations that do not condense into static short-range order. If a system displays static short-range magnetic order, the distribution of internal hyperfine fields will result in broadening of the NMR spectrum. In the present case, although the ${}^{31}$P NMR spectrum below $T_N$ is broadened by approximately an order of magnitude (see Fig.~\ref{fig:fig_9_FWHM_vs_temp}), no increased line broadening is observed above $T_N$ for either $H_0 \parallel c^*$ or $H_0 \perp c^*$. This apparent contradiction motivated a careful investigation of other known causes of the $K$--$\chi$ anomaly. These include crystal field depopulation and heavy-fermion behavior/Kondo physics.

Heavy-fermion behavior has been the topic of many previous NMR studies, and several interpretations of the Knight shift anomaly exist~\cite{Curro_2009_NMR_heavy_fermions}. No conclusive microscopic model currently exists, however a successful phenomenological model---typically referred to as the two fluid model---exists~\cite{Yang_2012_emergent}, which invokes the concept of Kondo screening, where the conduction electrons hybridize with and screen local moments. The result can be modeled by a conduction electron fluid and a heavy (hybridized) electron fluid. The hyperfine coupling is normally the proportionality constant between the bulk susceptibility and the Knight shift, however, if the hyperfine couplings to the local moments and the conduction elections are not equal and these systems interact, then it is possible for the scaling to break down~\cite{Shirer_2012_long_range}. However, this mechanism can be ruled out based on the fact that Ni$_2$P$_2$S$_6$ is a good semiconductor with an energy gap of $1.59 \pm 0.05$~eV~\cite{Foot_1980_TmPS3_optic_elec_props}, and therefore does not host the required conduction elections.

The second possible mechanism of breakdown in $K$--$\chi$ scaling is due to depopulation of excited crystal field energy levels. The picture here is that electrons in different crystal field levels have different hyperfine coupling, so upon decreasing temperature the higher energy levels will become depopulated and change the overall observed hyperfine coupling of the system~\cite{Ohama_1995_CeCu2Si2_nmr_kchi}. This scenario can be checked theoretically by employing quantum chemistry calculations to determine the crystal field levels. Our calculations (discussed in detail in Section~\ref{sec:quantum_chemistry}) show that the splitting between the Ni ground state crystal field triplet ($A_{2g}$) and the first excited triplet state ($T_{2g}$) is between 0.84 and 1.05\,eV. This is more than an order of magnitude too large to explain the observed anomaly temperature of approximately 275\,K, which corresponds to 0.024\,eV.

After ruling out these possibilities, we conclude that the $K$--$\chi$ anomaly is related to the onset of short-range quasi-2D correlations above $T_N$. This conclusion is made based on the proximity of the anomaly temperature $T^* \sim 275$\,K---marking the approximate onset of deviation from linearity in the ${}^{31}K$--$\chi$ plots (see Figs.~\ref{fig:fig_5_K_and_chi_vs_temp} and \ref{fig:fig_6_K_vs_chi_layout})---to the maximum in the susceptiblity $T_{\chi_\mathrm{max}} = 262 \pm 5$\,K. As mentioned above, this maximum in $\chi$ is a well-known consequence of reduced dimensionality on magnetic correlations. The minimum in ${}^{31}$K occurs at a temperature $T_{K_\mathrm{min}} \sim 240$\,K, just below the maximum in $\chi$, indicating that the local susceptibility is less impacted by reduced dimensionality in comparison to the bulk susceptibility. However, the lack of spectral broadening and conservation of spectral weight above $T_N$ indicate that, either the correlations do not fully condense into static short-range order, or the fraction of nuclei that experience static short-range order is smaller than the uncertainty in spectral weight. It is also possible that stacking faults play a roll by producing a distribution of interplane couplings, which in turn affect the local susceptibility differently from the global susceptibility.

This is somewhat different from the other quasi-2D magnets with known NMR shift anomalies. For example, while there was no mention of spectral broadening of the ${}^{61}$Ni NMR in K$_2$NiF$_4$, there was significant ${}^{19}$F NMR broadening in a separate study by Maarschall et al.\ \cite{Maarschall_1969_K2NiF4_NMR}. Sarkar et al.\ also find that the ${}^{23}$Na spectrum broaden continuously with decreasing temperature~\cite{Sarkar_2019_Na2IrO3_disorder}. Broadening of the NMR spectrum is expected in the presence of static short-range order, and is a result of a distribution of hyperfine fields produced by the short-range ordered moments coupling to the nuclei via off-diagonal terms in the hyperfine coupling tensor. If only a small fraction of the nuclei in the sample---up to roughly 5\,\%, based on the uncertainty in our spectral weight measurements---experience short range order, then this would suppress the bulk susceptibility, but be invisible to NMR.

Finally, it seems that a key ingredient to observing a shift anomaly in a quasi-2D system is a relatively high ordering temperature. For example, there exist several cases of quasi-2D compounds that show a similar hump in the $\chi$ and $K$, but no breakdown in scaling~\cite{Nath_2009_Pb2VOPO42_NMR, Ranjith_2015_Li2CuW2O8_NMR, Bossoni_2011_SrZnVOPO42}. The degree of frustration may also play some role, but this is beyond the scope of this work.

Another key finding of our study is the observation of appreciable transferred hyperfine coupling between the ${}^{31}$P nuclei and the surrounding electron magnetic moments on the Ni sites. Considering that the magnitude of the shift is quite small in general, we must first consider the direct dipolar hyperfine coupling mechanism. We calculate the dipolar hyperfine coupling tensor at the ${}^{31}$P site via a lattice sum method~\cite{Grafe_2017_LiCuSbO4_spin_chain} over a radius of 600~{\AA}, based on lattice parameters from X-Ray diffraction~\cite{Fragnaud_1993_NiPS3_lattice_params} and magnetic structure/ordered moment from neutron scattering~\cite{Wildes_2015_NiPS3_neutron_magstruc}. The calculated hyperfine tensor is given by
\begin{equation}
    \label{eqn:calc_dip_hyp_tensor}
    \mathcal{A}_\mathrm{dip} = 
    \begin{bmatrix}
        A_{xx} & A_{xy} & A_{xz} \\
        A_{yx} & A_{yy} & A_{yz} \\
        A_{zx} & A_{zy} & A_{zz} \\
    \end{bmatrix}
    =
    \begin{bmatrix*}[r]
         0.058 & 0.009 & -0.003 \\
         0.009 & 0.070 &  0.003 \\
        -0.003 & 0.003 & -0.129 \\
    \end{bmatrix*},
\end{equation} 
where all values are in T/$\mu_B$. By subtracting the calculated dipolar hyperfine couplings from the measured experimental values we can estimate the transferred hyperfine coupling $A_\mathrm{tr}$. These quantities are shown in Table~\ref{tab:tr_hf_coupling}. It is important to note that the values in the table were rounded to the first decimal place of the uncertainty after the calculation was performed.

We find that the in-plane component of $A_\mathrm{tr}$ is appreciably large at $-0.5 \pm 0.2$\,T/$\mu_B$, indicating that there is hybridization of Ni orbitals with those of the P (mediated by the interstitial S, due to the P-S covalent bonding). This finding is consistent with Ni--S--Ni super-exchange and  Ni--S--S--Ni super-super-exchange, which were proposed as driving mechanisms for the large values of nearest neighbor and third nearest neighbor exchange coupling parameters, $J_1$ and $J_3$ respectively~\cite{Lancon_2018_NiPS3_INS}. A related caveat is that some spin polarization may exist on the sulfur sites: DFT+$U_\mathrm{eff}$ calculations from Kim et al.\ \cite{Kim_2018_NiPS3_correlations}, find an ordered moment of 0.15~$\mu_B$ on the S(2) site, which points in the same direction as the neighboring Ni moments. This spin polarization could then be transferred via S--P orbital hybridization to generate a transferred hyperfine field at the ${}^{31}$P site. 

\begin{table}
\begin{ruledtabular}
\begin{tabular}{cccc}
           & {$A_\mathrm{tot}$ (T/$\mu_B$)} & {$A_\mathrm{dip}$ (T/$\mu_B$)} & {$A_\mathrm{tr}$ (T/$\mu_B$)} \\[0.25ex]
\hline \\[-1.5ex]
$A_{c^*}$  & $-0.2 \pm 0.1$                 & $-0.129$                       & $ 0.1 \pm 0.1$              \\[0.5ex]
$A_{ab}$   & $-0.5 \pm 0.2$                 & $ 0.064$                       & $ 0.5 \pm 0.2$                \\[0.5ex]
\end{tabular}
\end{ruledtabular}
\caption{\label{tab:tr_hf_coupling}Experimentally observed total hypefine coupling $A_\mathrm{tot}$, calculated dipolar coupling $A_\mathrm{dip}$, and resultant transferred hyperfine coupling $A_\mathrm{tr} = A_\mathrm{tot} - A_\mathrm{dip}$. $A_\mathrm{tot}$ are the high temperature values from Table~\ref{tab:hf_coupling_exp}. $A_{ab,\mathrm{dip}}$ is an average of $A_{xx}$ and $A_{yy}$ from Eqn.~\ref{eqn:calc_dip_hyp_tensor}.}
\end{table}

We now change the focus of our discussion to the ${}^{31}$P $T_1^{-1}$ measurements, which are sensitive to spin fluctuations. We observe no indication of critical enhancement of spin fluctuations above $T_N$. Instead, a broad maximum in $(T_1T)^{-1}$ vs temperature is observed (see Fig.~\ref{fig:fig_8_T1}), which coincides qualitatively with the broad maximum in the magnetic susceptibility and the minimum in the NMR shift. $T_1^{-1}$ is found to be nearly isotropic, with slightly faster relaxation for $H_0 \parallel c^*$. This is consistent with our measurements of the spectral splitting as a function of angle as discussed below in Section~\ref{subsec:magnetic_state}, which reveal that the internal field at the ${}^{31}$P site lies dominantly in the basal plane.

Power-law fits of $T_1^{-1}$ vs temperature (see Section~\ref{subsec:spin--lattice_relaxation}) yield $T_1^{-1} \propto T^5$, which is indicative of relaxation dominated by three-magnon scattering~\cite{Beeman_1968_nuclear}. This relation should hold as long as the temperature is larger than the spin gap in the magnon dispersion. Our data follow a $T^5$ dependence down to approximately 40\,K, whereas the spin gap measured by inelastic neutron scattering at the Brillouin zone center---and calculated to be approximately the same at the zone edge---is on the order of 7\,meV (80\,K). At lower temperatures, instead of gap-like activated behavior, $T_1^{-1}$ approaches a constant value, indicative of an additional relaxation channel.

Previous measurements in several $M_2$P$_2X_6$ compounds found field dependence of the magnetic transitions based on ${}^{31}$P $T_1^{-1}$~\cite{Ziolo_1988_31P_NMR_MPX3, Torre_1989_MPX3_31P_NMR}. Magnetization measurements of our high-quality single crystals showed no such dependence (see Fig.~\ref{fig:fig_2_TN_vs_field}), which is also in agreement with the $M/H$ measurements of Wildes et al.\ \cite{Wildes_2015_NiPS3_neutron_magstruc}. Even so, we investigated the field dependence of $T_1^{-1}$ for $H_0 \perp c^*$ just above $T_N$ at $T=165$\,K, where we would expect to see a strong field dependence based on the above mentioned data from the literature. However, as shown in Fig~\ref{fig:fig_8_T1}(b) as $(T_1T)^{-1}$ vs field $H_0$, we find little to no field dependence to within the standard error of our measurements. Even if the slight suppression at the two lowest fields is real, this is a much smaller effect than previously observed. Based on our magnetization and $T_1$ measurements, we conclude that the previously observed field dependence is absent in our crystals. We speculate that the previously observed field dependence~\cite{Ziolo_1988_31P_NMR_MPX3, Torre_1989_MPX3_31P_NMR} may be related to sample quality issues, especially considering disagreement with data from simulated powder patterns based on our single-crystal data (see Appendix~\ref{sec:31P_powder_patterns} for further details).

\begin{figure} 
    \includegraphics[trim=0cm 0cm 0cm 0cm, clip=true, width=\linewidth]{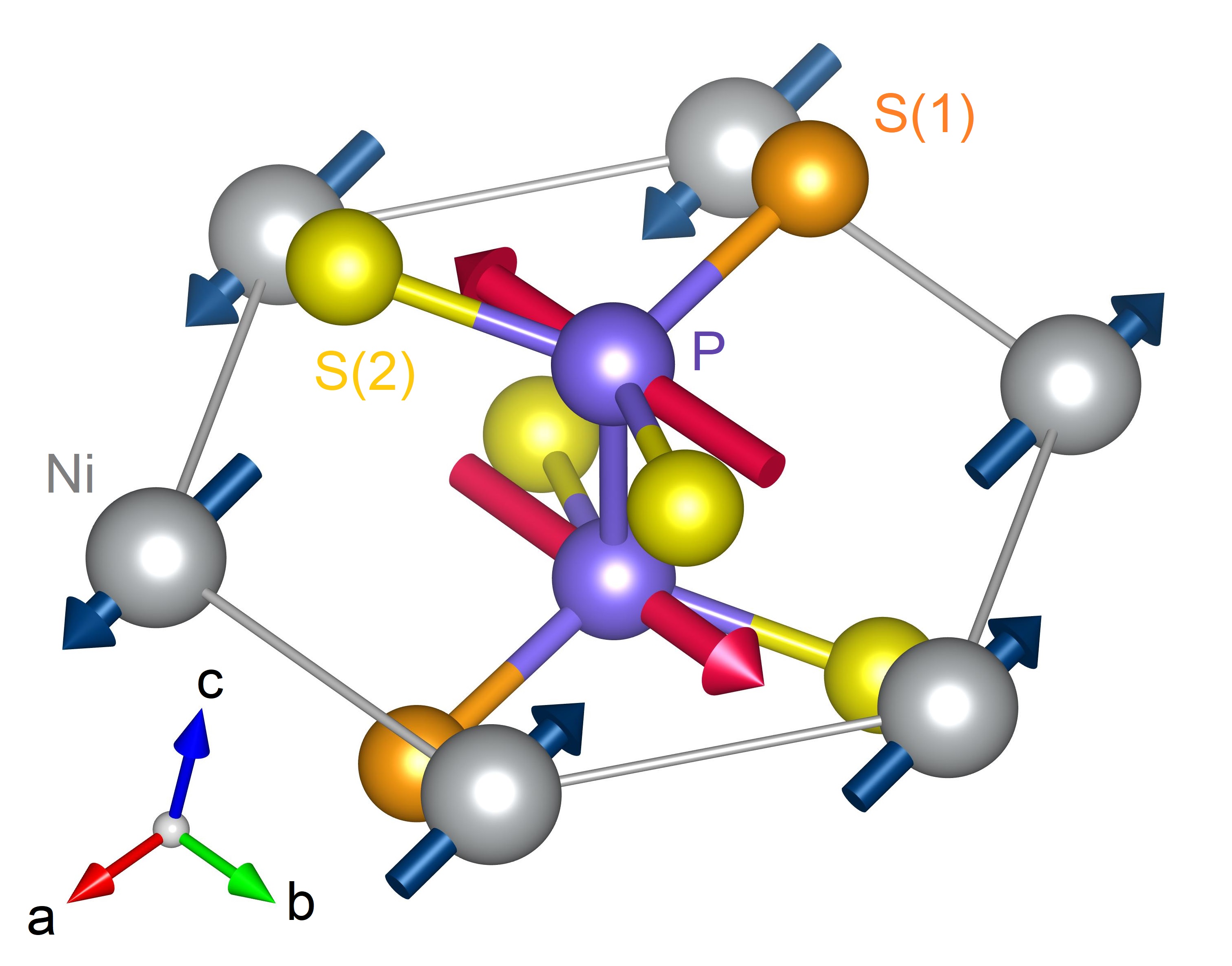}
    {\caption{\label{fig:fig_11_NiPS3_AFM_hyp}Local environment of the ${}^{31}$P sites in the antiferromagnetically ordered state~\cite{Fragnaud_1993_NiPS3_lattice_params, Momma_2011_VESTA3}. Navy blue vectors represent ordered moments of 1.05\,$\mu_B$ at the Ni sites (from neutron diffraction measurements~\cite{Wildes_2015_NiPS3_neutron_magstruc}). Dark pink vectors represent the internal hyperfine fields at the P sites, calculated by a dipolar lattice sum as discussed in the text.}}
\end{figure}

Turning to our magnetic state spectral measurements, we extrapolate the measured internal field---$H_\mathrm{int}^\mathrm{exp}(T = 150 K) = 0.0105$\,T extracted from the angular dependent spectra---to zero temperature, based on the fit to the temperature dependence of the FWHM, to be $H_\mathrm{int}^\mathrm{exp}(T=0~\mathrm{K}) = 0.027$\,T. Note that the 150\,K simulations are not a least-squares fit, and therefore do not have a well-characterized uncertainty. This value is larger than the magnitude of the calculated dipolar internal hyperfine field (approximately 0.015\,T). The full internal field vectors are $H_\mathrm{int,above}^\mathrm{dip}$ = 0.0003~$\hat{a}$ -- 0.0156~$\hat{b}$ -- 0.0001~$\hat{c^*}$, and $H_\mathrm{int,below}^\mathrm{dip}$ = 0.0002~$\hat{a}$ + 0.0153~$\hat{b}$ + 0.0002~$\hat{c^*}$, where above/below indicates the unique magnetic P site above/below the Ni plane (see Fig.~\ref{fig:fig_11_NiPS3_AFM_hyp} for a full visualization). This discrepancy provides further evidence for an appreciable transferred hyperfine coupling, and therefore supports the conclusion that Ni--S--Ni super-exchange and Ni--S--S--Ni super-super-exchange are responsible for the large $J_1$ and $J_3$ exchange couplings~\cite{Lancon_2018_NiPS3_INS}.

The angular dependences of the NMR spectra in the magnetic state (see Section~\ref{subsec:magnetic_state} and Figure~\ref{fig:fig_10_magnetic_state_ang_dep}) are best described by stacking faults along the $c^*$-axis, which occur with $\pm 60$ degree misalignment with respect to the planes below. This effect is detectable by NMR after the in-plane symmetry is broken by the appearance of an internal hyperfine field in the magnetic state, as shown in Fig.~\ref{fig:fig_11_NiPS3_AFM_hyp}. This interpretation is backed up by our scXRD measurements (see Appendix~\ref{sec:elemental_comp_and_xrd}) as well as those of Goossens et al.\ \cite{Goossens_2011_NiPS3_stacking_faults}. Neutron scattering results~\cite{Wildes_2015_NiPS3_neutron_magstruc, Lancon_2018_NiPS3_INS} also indicate that out-of-plane stacking faults are prevalent in Ni$_2$P$_2$S$_6$ and that the magnetic order is coupled to the lattice. While NMR is not quantitatively sensitive to the domain sizes in this case, the difference in spectral weight for the three unique pairs of magnetically split resonances indicates that the domains likely have thicknesses larger than a few unit cells. M\"{o}{\ss}bauer spectroscopy measurements in Fe$_2$P$_2$S$_6$ provide evidence for magnetic microdomains, which are most likely associated with stacking faults as well~\cite{Jernberg_1984_FePS3_Moessbauer}. Stacking faults are also known to be present in the related In$_2$Ge$_2$Te$_6$~\cite{Lefevre_2017_In2Ge2Te6}.

Our NMR spectral measurements as a function of in-plane angle $\phi$ in the paramagnetic state also do not agree with the anti-site disorder picture (Ni trading places with a P dimer). Anti-site disorder would break the in-plane local environment symmetry and result in $K_a \neq K_b$ for a large fraction of the observed nuclei, yet after accounting for the P--P homonuclear dipolar coupling (the Pake doublet), we find $K_a = K_b$. The disorder picture, at the levels suggested by scXRD refinement, would likely result in significant broadening of the NMR spectra in magnetic state. Such a high fraction of anti-site disorder would also likely affect the magnetic properties of the system, particularly with respect to suppression of $T_N$.

To conclude, we investigated high quality crystals of the quasi-2D van der Waals antiferromagnet Ni$_2$P$_2$S$_6$ via NMR, magnetic susceptibility, scXRD, and quantum chemistry calculations. We have shown that NMR is sensitive to quasi-2D magnetic correlations via an anomalous breakdown in the scaling ${}^{31}K$ vs $\chi$, possibly also affected by a distribution of stacking-fault-induced interplane couplings. Our quantum chemistry calculations show that the source of this breakdown is unrelated to crystal field depopulation effects. We find an appreciable in-plane transferred hyperfine coupling, consistent with super- and super-super-exchange coupling. Our $M/H$ and $T_1$ measurements show that $T_N$ is field independent. $T_1^{-1}$ measurements also provide evidence for three-magnon relaxation in the magnetic state. Our magnetic state NMR spectra provide good evidence for 60 degree rotation of stacking-fault-induced magnetic domains. Our work motivates future experiments in related $M_2$P$_2X_6$ systems, as well as other quasi-2D van der Waals magnets, to develop a microscopic description of the $K$--$\chi$ anomaly.

\begin{acknowledgments}
The authors would like to acknowledge helpful discussions with, and give our thanks to H.\ Yasuoka, G.\ Bastien, G.\ Shipunov, P.\ Fritsch, C.\ He{\ss}, N.\ J.\ Curro, and P.\ Lepucki. A.\ P.\ Dioguardi was supported by Deutsche Forschungsgemeinschaft (DFG) Grant No.\ DI2538/1-1. S.\ Aswartham acknowledges financial support from DFG Grant No.\ AS 523/4-1. S.\ Aswartham and S.\ Selter acknowledge financial support from GRK-1621 graduate academy of the DFG. M.\ Sturza acknowledges financial support from DFG Grant No.\ STU 695/1-1. R.\ Murugesan, M.\ S.\ Eldeeb, and L.\ Hozoi thank U.\ Nitzsche for technical support.
\end{acknowledgments}

\appendix

\begin{figure}
    \includegraphics[trim=0cm 0cm 0cm 0cm, clip=true, width=\linewidth]{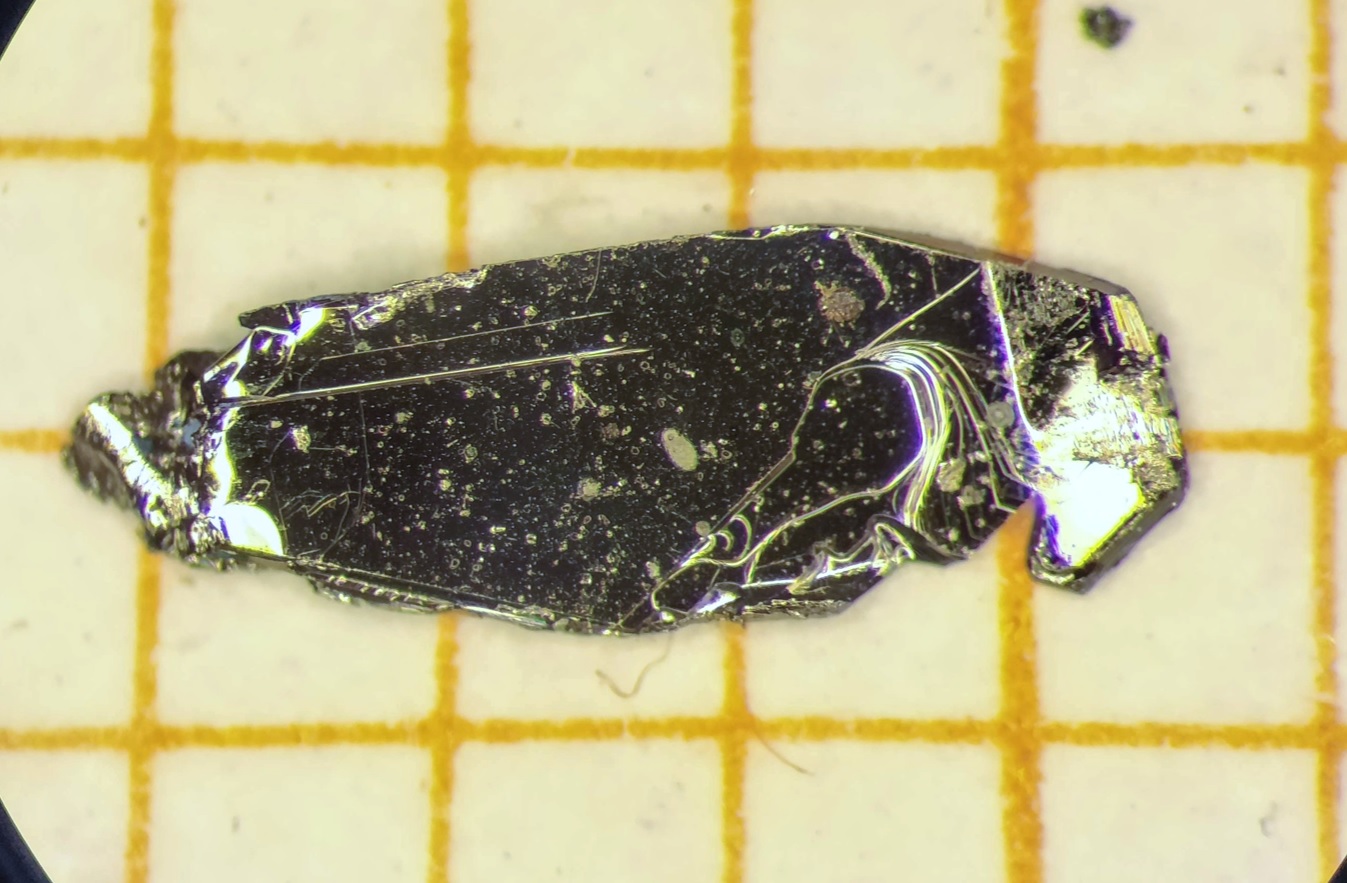}
    \includegraphics[trim=0cm 0cm 0cm 0cm, clip=true, width=\linewidth]{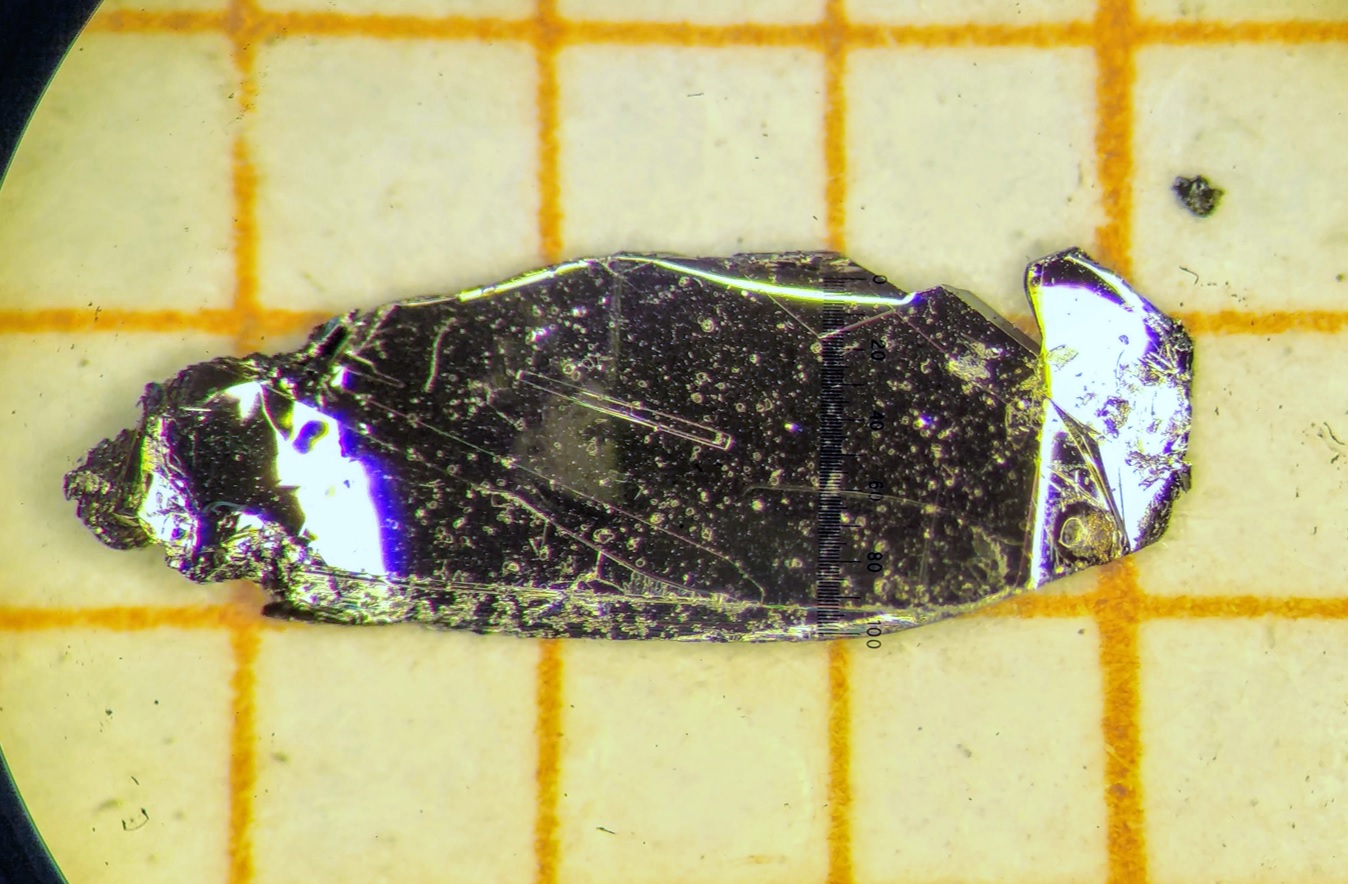}
    {\caption{\label{fig:fig_12_xtal_a_photo}Photographic images of the front and back of a single crystalline sample of Ni$_2$P$_2$S$_6$ (NMR crystal A). The background grid has divisions of 1\,mm.}}
\end{figure}

\section{Elemental Composition \& Structural Characterization}
\label{sec:elemental_comp_and_xrd}

scXRD was performed at room temperature on a Bruker\,X8\,Apex2\,CCD4K diffractometer with Mo-K$_{\alpha}$ radiation. The data collection consists of large $\Omega$ and $\phi$ scans of the reciprocal space. The frames were integrated with the Bruker SAINT software package~\footnote{SAINT, Bruker AXS Inc., Madison, Wisconsin, USA (2004)} using a narrow-frame algorithm in APEX2~\footnote{APEX2, Bruker AXS Inc., Madison, Wisconsin, USA (2004)}. The data were corrected for absorption effects using a semiempirical method based on redundancy with the SADABS program~\footnote{G. M. Sheldrick, SADABS, Program for Empirical Absorption Correction of Area Detector Data, University of G{\"{o}}ttingen, Germany (1996)}, developed for scaling and absorption corrections of area detector data. The space group determination, structural determination and refinement were performed using charge flipping with the Superflip algorithm~\cite{Palatinus_2007_superflip} within Jana2006~\cite{Petricek_2014_JANA2006} and SHELXL~\cite{Sheldrick_2008_SHELX}. The parameters for data collection and the details of the structure refinement are given in Table~\ref{tab:scXRD_summary}. A ZEISS EVO MA 10 SEM with a BSE detector was used for microscopic crystal images with chemical contrast. EDX was measured at an accelerating voltage of 30\,kV using a energy dispersive X-ray analyzer mounted to a the SEM.
Fig.~\ref{fig:fig_12_xtal_a_photo} shows a photograph of Ni$_2$P$_2$S$_6$ crystal A used for NMR. The background shows a millimeter grid for scale. Good examples of the crystalline facets can be seen at the top right of the photograph. SEM(BSE) images of the crystals (not shown) display uniform contrast, indicating a homogeneous distribution of elements. The mean elemental composition of our crystals was found to be $19.9 \pm 0.6$\,at-\%\,Ni, $20.4 \pm 0.1$\,at-\%\,P and $59.7 \pm 0.6$\,at-\%\,S by EDX measurements on several spots on different crystals. We note, that the systematic uncertainty of EDX is in the range of approximately $\pm 3$\% even on flat sample surfaces.

\begin{figure}
    \includegraphics[trim=0cm 0cm 0cm 0cm, clip=true, width=\linewidth]{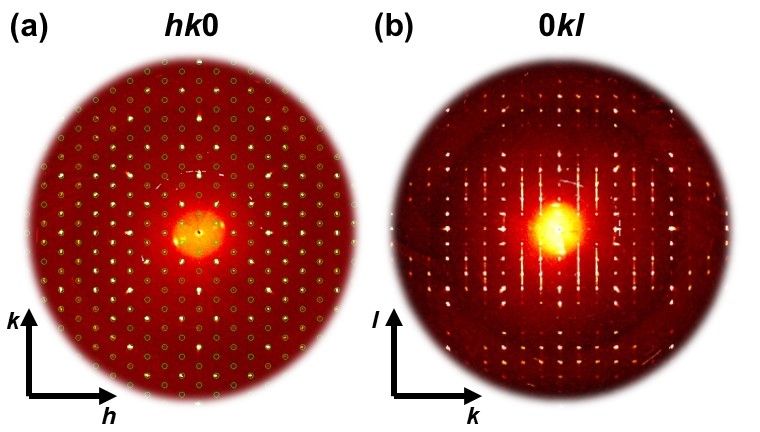}
    {\caption{\label{fig:fig_13_rec_space_layers}Cuts through reciprocal space from scXRD showing the (a) $hk0$ and (b) $0kl$ planes. Circles in (a) indicate the expected reflection positions according to the structural model.}}
\end{figure}

All diffraction spots in reciprocal space of the measured crystal could be indexed by the reported monoclinic space group for Ni$_2$P$_2$S$_6$ of $C 1 2/m 1$ (No. 12)~\cite{Ouvrard_1985_MPS3_sx_XRD, Wildes_2015_NiPS3_neutron_magstruc} as indicated by small circles in Fig.~\ref{fig:fig_13_rec_space_layers}(a). Structural refinement based on the scXRD data resulted in a structural model for our crystal in good agreement with the reported crystal structure.

\begin{table*}
 \begin{tabular}{c c}
 \hline \hline
 Empirical Formula & Ni$_2$P$_2$S$_6$ \\
 \hline
 Formula Weight & 371.72 \\
 Temperature & $293(2)$\,K \\
 Wavelength & 0.71073\,\AA \\
 Crystal System & Monoclinic \\
 Space Group & $C$2/$m$ \\
 Unit Cell Dimensions & $a = 5.8165(7)$\,\AA \\
 & $b = 10.0737(12)$\,\AA \\
 & $c = 6.6213(8)$\,\AA \\
 & $\beta = 107.110(6)^\circ$ \\
 Volume & $370.80(8)$\,\AA$^3$ \\
 Z & 4 \\
 Density(calculated) & 3.329 g/cm$^3$ \\
 Absorption Coefficient & 7.094 mm$^{-1}$ \\
 F(000) & 364 \\
 $\theta$ Range for Data Collection & 3.219-43.225$^\circ$ \\
 Index Ranges & $-11 \leq h \leq 11$, $-19 \leq k \leq 19$ \\
 & $-12 \leq l \leq 12$ \\
 Reflections Collected & 19885 \\
 Independent Reflections & 1438 (R$_{int} = 0.0445$) \\
 Completness of $\theta = 26.64^\circ$ & 100\% \\
 Refinement Method & Full-matrix least square on F$^2$ \\
 Data / Restraints / Parameters & 1438 / 0 / 45 \\
 Goodness-Of-Fit & 1.080 \\
 Final R Indices [$>2\sigma(I)$] & R$_{obs} = 0.0222$, wR$_{obs} = 0.0499$ \\
 R Indices [all data]$^a$ & R$_{all} = 0.0292$, wR$_{all} = 0.0526$ \\
 Extinction Coefficient & $0.0047(8)$ \\
 Largest Diff. Peak and Hole & 1.823 and -0.581 e$\cdot$\AA$^{-3}$ \\
 \hline \hline
 \multicolumn{2}{l}{$^aR = \Sigma||F_O|-|F_C||/\Sigma|F_O|$, $wR = \lbrace\Sigma[w(|F_O|^2-|F_C|^2)^2]/\Sigma[w(|F_O|^4)]\rbrace^{1/2}$ and}\\
 \multicolumn{2}{l}{$w = 1/[\sigma^2(F_O^2)+(0.0496P)^2+0.8710P]$ where $P = (F_O^2+2F_C^2)/3$}\\
 \end{tabular}
\caption{Summary of crystallographic data and structure refinement for Ni$_2$P$_2$S$_6$ at $293(2)$\,K.}
\label{tab:scXRD_summary}
\end{table*}

\begin{table*}
\begin{tabular}{ccccccc}
\hline \hline
Label & Wyckoff & x & y & z & Occupancy & ${\textrm{U}_{eq}}^a$ \\
\hline
Ni(1) & 4g & 0 & 3331(1) & 0 & 0.963(2) & 10(1) \\
P(1) & 4i & 576(2) & 0 & 1699(1) & 0.914(4) & 9(1) \\
S(1) & 4i & 7422(1) & 0 & 2432(1) & 1 & 9(1) \\
S(2) & 8j & 2516(1) & 1698(1) & 2434(1) & 1 & 9(1) \\
Ni(2) & 2a & 0 & 0 & 0 & 0.077(3) & 42(2) \\
P(2) & 8j & 510(18) & 3335(6) & 1530(30) & 0.029(3) & 102(5) \\
\hline \hline
\multicolumn{7}{l}{$^a$U$_{eq}$ is defined as one third of the trace of the orthogonalized U$_{ij}$ tensor.}
\end{tabular}
\caption{Atomic coordinates ($\times 10^4$) and equivalent isotropic displacement parameters ${\textrm{U}_{eq}}$ (\AA$^2 \times 10^3$) of Ni$_2$P$_2$S$_6$ at $293(2)$\,K with estimated standard deviations in parentheses.}
\label{tab:scXRD_Coordinates}
\end{table*}

\begin{figure*}
    \includegraphics[trim=0cm 0cm 0cm 0cm, clip=true, width=\linewidth]{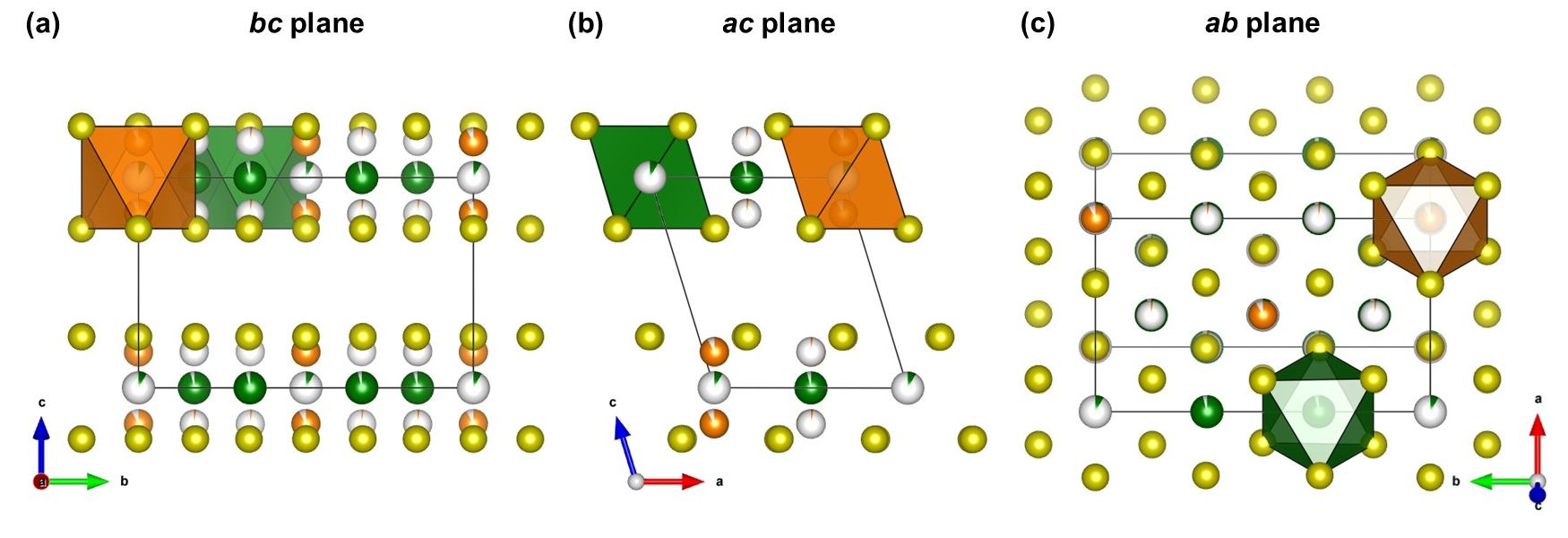}
    {\caption{\label{fig:fig_14_cryst_structure}Structural model of Ni$_2$P$_2$S$_6$ obtained from scXRD. Green balls represent Ni, orange balls represent P, and yellow balls represent S. (a) View along the $a$ direction perpendicular to the $bc$ plane, (b) view along the $b$ direction perpendicular to the $ac$ plane, and (c) view along the $c^*$ direction perpendicular to the $ab$ plane.}}
\end{figure*}

As shown by Ouvrard et al.\ \cite{Ouvrard_1985_MPS3_sx_XRD} and Wildes et al.\ \cite{Wildes_2015_NiPS3_neutron_magstruc}, introducing site disorder between the majority 4\textit{g} and the minority 2\textit{a} sites for Ni and between the majority 4\textit{i} and minority 8\textit{j} sites for P improves the agreement between structural model and experimental diffraction data. In our model, the best agreement with experiment is obtained for approximately 4\,\% of Ni atoms and 6\,\% of P atoms on the respective minority sites. Additionally, the structural refinement indicates a small amount of vacancies on the P sites, which results in a refined formula of Ni$_2$P$_{1.94}$S$_6$. The obtained structural model is shown in Fig.~\ref{fig:fig_14_cryst_structure} with the corresponding atomic coordinates given in Table~\ref{tab:scXRD_Coordinates}.

Additionally, the diffraction pattern shows a significant broadening of reflections in the l direction (equivalent to the $c$* direction in real space) (Fig.~\ref{fig:fig_13_rec_space_layers}(b)) of the 0kl layer. This broadening is strongly indicative of a high concentration of stacking faults, which is a well known defect in layered van der Waals compounds and was observed in Ni$_2$P$_2$S$_6$ by Goossens~et~al.~\cite{Goossens_2011_NiPS3_stacking_faults} and Lan\c{c}on~et~al.~\cite{Lancon_2018_NiPS3_INS}. As discussed in both aforementioned works, it is likely that the displaced electron density resulting from these stacking faults is misinterpreted in the structural solution and falsely leads to a crystal structure model involving site disorder. Consequently, the scXRD analysis yields a crystal structure model for our Ni$_2$P$_2$S$_6$ crystal that is in good agreement with the structure of the $M_2$P$_2$S$_6$ family in the space group $C 1 2/m 1$ and indicates a high concentration of defects in the form of stacking faults. The existence of site disorder in our crystal cannot be determined unambiguously from scXRD.

\section{Spectral splitting due to homonuclear dipolar coupling}
\label{sec:pake_doublet}

\begin{figure}
    \includegraphics[trim=0cm 0cm 0cm 0cm, clip=true, width=\linewidth]{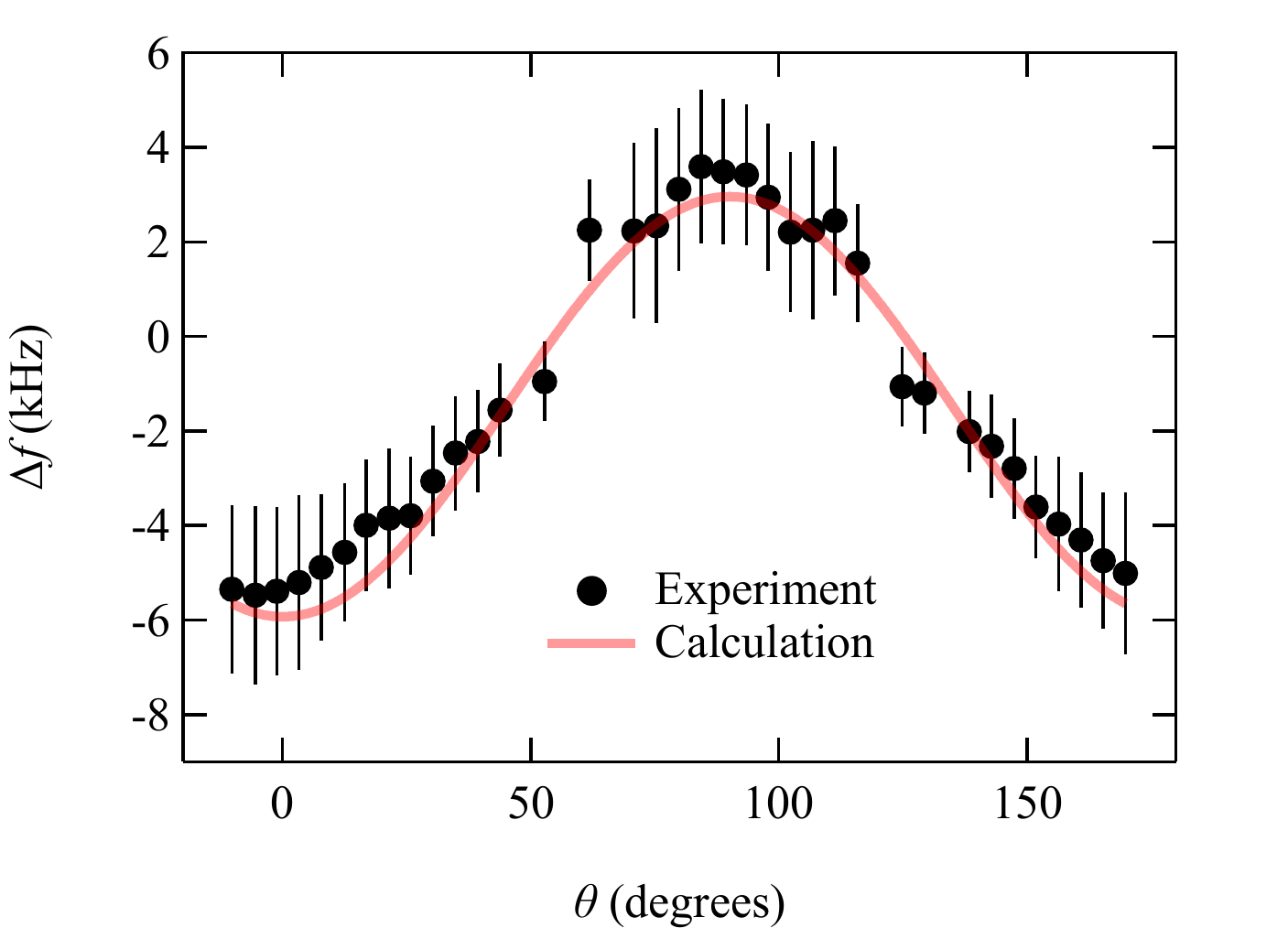}
    {\caption{\label{fig:fig_15_pake_doublet_splitting}Angular dependence of the splitting between the two observed resonances $\Delta f$ in crystal A (markers) and the calculated $\Delta f$ (red curve) for a P--P separation of 2.1534\,{\AA} (extracted from the crystal structure refined from scXRD).}}
\end{figure}

The angular dependence of the splitting of the NMR spectrum agrees well with the expected behavior for dipolar coupling between the two ${}^{31}$P nuclear spins in the P--P dimer (see Fig.~\ref{fig:fig_15_pake_doublet_splitting}). This is a well-known phenomenon refered to as a Pake doublet, that was also suggested by Berthier et al.\ to explain the broadening of their powder pattern~\cite{Berthier_1978_MPS3_NMR_Li, Pake_1948_pake_doublet}. We note that the splitting observed here is far smaller than would be required to  explain the line broadening in the literature. This homonuclear dipole-dipole interaction term commutes with the other terms in the total nuclear spin Hamiltonian, and therefore can simply be subtracted off to access the relevant electron-nuclear interactions that we wish to probe. In practice, our data reduction is achieved by finding the center of gravity of the spectrum (average of the two resonance frequencies). The angular dependence of the splitting in the Pake doublet is given by
\begin{equation}
\label{eqn:pake_doublet_splitting}
\Delta f (\theta) = f_1 - f_2 = \frac{3}{2}\frac{\mu_0}{4 \pi}\frac{\hslash \gamma^2}{r^3}\left(1 - 3 \cos^2{\theta}\right),
\end{equation}
where $\mu_0$ is the permeability of free space, $\hslash$ is Planck's constant divided by $2\pi$, $\gamma$ is the gyromagnetic ratio, and $r$ is the distance between the nuclear spins ($r_\mathrm{P-P} = 2.1534$\,{\AA}).

\section{Search for ${}^{33}$S and ${}^{61}$Ni NMR}
\label{sec:33S_and_61Ni_NMR}

${}^{33}$S and ${}^{61}$Ni NMR measurements were also attempted, but without success. This is likely a consequence of the low natural abundance of the NMR-active isotopes ${}^{61}$Ni (1.1399\%) and ${}^{33}$S (0.76\%)~\cite{Harris_2001_NMR_isotopes}. Furthermore, the on-site magnetic moment of Ni likely contributes to the lack of signal via a combination of large shift, linewidth, and fast relaxation rates. Additionally, previous calculations (DFT+$U_\mathrm{eff}$) also found some spin density (0.15~$\mu_B$) on the S sites closest to the zig-zag Ni chains, which may contribute to the lack of ${}^{33}$S signal~\cite{Kim_2018_NiPS3_correlations}.

\section{Corrections due to macroscopic magnetism}
\label{sec:mac_mag_correct}

In the case of samples with nonspherical geometry, one must take into account the corrections due to classical magnetism. The two contributions are demagnetization field $\mathbf{h}_D$ (due to the sample's shape), and the Lorentz field $\mathbf{h}_L$ (due to the uniformly magnetized bulk outside of a sphere surrounding the nucleus)~\cite{Zimmerman_1957_stand_NMR_spec}. Therefore, the total macroscopic magnetic field within the sample is $\mathbf{H}_i = \mathbf{H}_0 + \mathbf{h}_D + \mathbf{h}_L$. The Lorentz sphere contribution is given by,
\begin{equation}
    \mathbf{h}_L = \frac{4}{3}\pi \mathbf{M},
\end{equation}
where $\mathbf{M}$ is the sample's magnetization. The sample's shape-dependent demagnetization field is given by 
\begin{equation}
    \mathbf{h}_D = -\mathbb{D} \cdot \mathbf{M},
\end{equation}
where $\mathbb{D}$ is the demagnetization factor tensor.

We calculated the demagnetization tensor elements based on the approach of Osborn~\cite{Osborn_1945_demag_factors}, assuming an ellipsoidal sample. The dimensions of the ellipsoid are taken from crystal A, shown in Fig.~\ref{fig:fig_12_xtal_a_photo}, with $a=3.631$\,mm, $b=1.241$\,mm, and $c=0.092$\,mm, associated with $L$, $M$, and $N$, respectively. The resulting demagnetization factors are $L/4\pi = 0.013$, $M/4\pi = 0.064$, and $N/4\pi = 0.923$. We note that the dimensions of crystal B were nearly the same as crystal A. Furthermore, the uncorrected angular dependent shifts were identical to within the experimental uncertainty (see Fig.~\ref{fig:fig_7_K_normal_ang_dep} in the main text). 

\begin{figure} 
    \includegraphics[trim=0cm 2cm 0cm 0cm, clip=true, width=\linewidth]{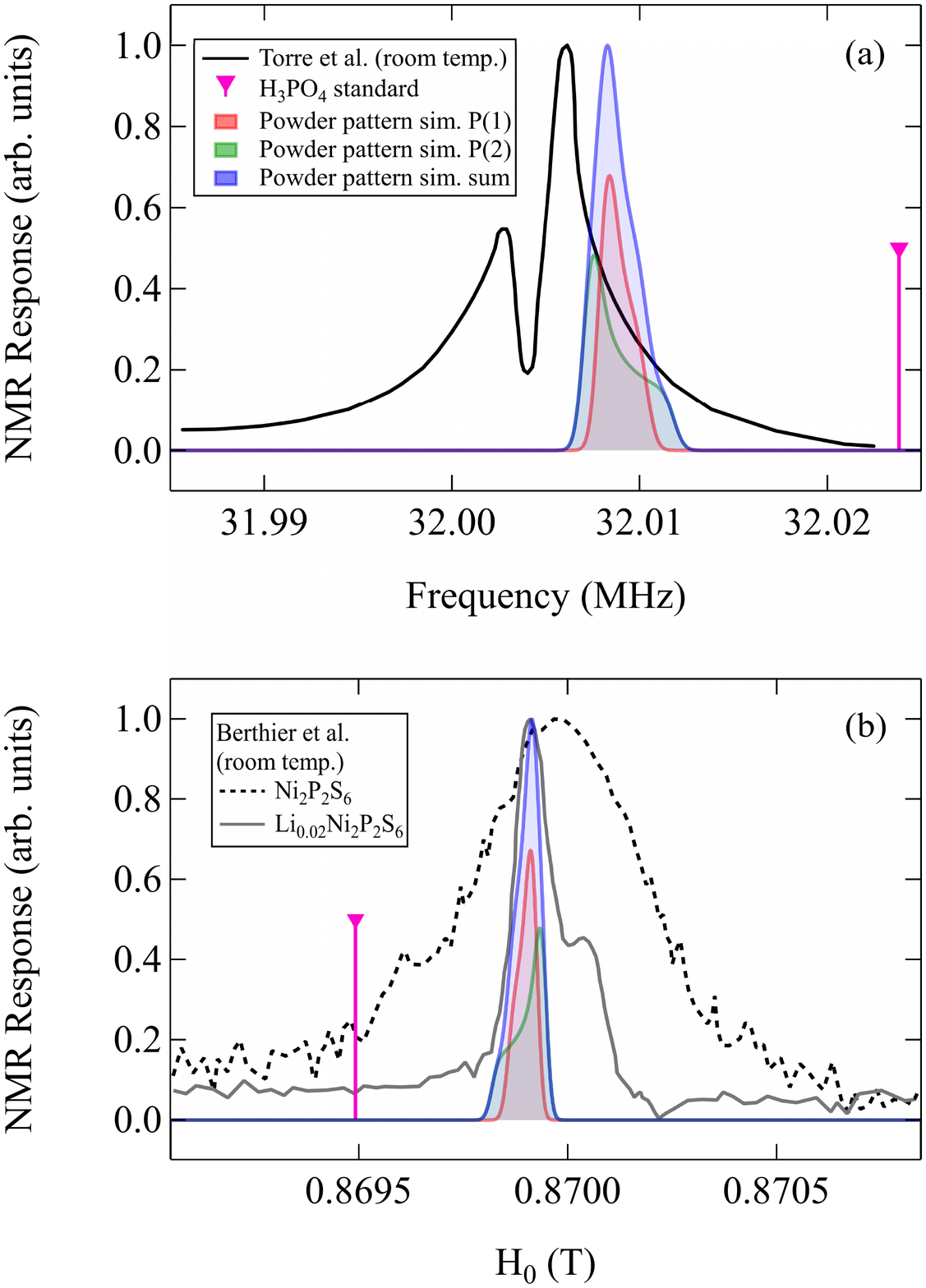}
    {\caption{\label{fig:fig_16_ppspec_comparison}(a) Comparison of the frequency-swept powder pattern (calculated from our single crystal angular dependent measurements) with that of Torre et al.\ \cite{Torre_1989_MPX3_31P_NMR}. (b) Comparison of the calculated field-swept powder pattern with data from Berthier et al.\ \cite{Berthier_1978_MPS3_NMR_Li}.}}
\end{figure}

To calculate the total macroscopic magnetism correction to the shift 
$K_d$ as a function of angle, we perform a rotation of the tensor expressions for $\mathbf{H}_d = \mathbf{h}_L + \mathbf{h}_D$, with respect to the external field $\mathbf{H}_0$ for out-of-plane and in-plane rotations. We then calculate the total shift due to macroscopic magnetism $K_d = |\mathbf{H}_i|/|\mathbf{H}_0| - 1$ for the cases of out-of-plane rotation ($\theta$ dependence) and in-plane rotation ($\phi$ dependence). The equations for the out-of-plane and in-plane rotation dependencies are given by Eqn.~\ref{eqn:out-of-plane_Kd} and Eqn.~\ref{eqn:in-plane_Kd}, respectively.

\begin{widetext}
\begin{align}
\label{eqn:out-of-plane_Kd}
K_d(\theta) &= \left(4 \pi^2 \chi_v^2 \sin^2{(2 \theta)} (M - N)^2 + \left(1+ \frac{2}{3} \pi \chi_v (3 \cos{(2 \theta)} (M - N) - 3 M - 3 N + 2)\right)^2\right)^\frac{1}{2} - 1 \\
\label{eqn:in-plane_Kd}
K_d(\phi) &= \left(4 \pi ^2 \chi_v^2 \sin^2{(2 \phi)} (L - M)^2 + \left(1 - \frac{2}{3} \pi  \chi_v (3 \cos{(2 \phi)} (L - M) + 3 L + 3 M - 2)\right)^2\right)^\frac{1}{2} - 1
\end{align}
\end{widetext}

The volume susceptibility $\chi_v = \chi (d/m)$ was calculated based on the measured molar susceptibility $\chi$, which was taken to be isotropic (the average value of the $M/H(T > T_N) \equiv \chi$ for $H \parallel c^*$ and $H \perp c^*$ were used for all corrections~\ref{fig:fig_1_MoverH_vs_temp}(b)). The molar mass $m=185.862$\,g/mol and sample density $d=3.325$\,g/cm$^3$ were taken from standard atomic weights and lattice parameters determined via scXRD, respectively. For the susceptibility itself, the demagnetization correction is small enough to be neglected. In the case of the NMR shift, there is an appreciable effect for both out-of-plane rotation and temperature dependent measurements. The in-plane angular dependence was not significantly affected outside of the experimental uncertainty, though the overall value was shifted.

\section{Calculated ${}^{31}$P NMR Powder Patterns}
\label{sec:31P_powder_patterns}

Initial measurements of powder samples by Berthier et al.\ found $K_\mathrm{iso} = -0.057(1)$\,\% at $T = 273$\,K~\cite{Berthier_1978_MPS3_NMR_Li}, but were unable to observe any clear asymmetry in their powder spectrum of pure Ni$_2$P$_2$S$_6$, and therefore report no value for $K_\mathrm{ax}$. In comparison, we find an average value (of the two Pake-doublet resonances) of $K_\mathrm{iso}^\mathrm{avg} = -0.04682 \pm 0.00009$\,K. Note that previous reports did not account for corrections due to macroscopic magnetism, and therefore the above value of $K_\mathrm{iso}^\mathrm{avg}$ is uncorrected. On the other hand, the powder spectrum of Torre et al.\ did show spectral splitting~\cite{Torre_1989_MPX3_31P_NMR}. 

We compare the calculated powder spectra to the spectra of Torre et al.\ and Berthier et al.\ in Fig.~\ref{fig:fig_16_ppspec_comparison}(a) and (b), respectively. We treat the two peaks, labeled P(1) and P(2), of the Pake doublet as unique site for these calculations. The spectral broadening of the powder pattern was applied via convolution with a Gaussian, scaled appropriately by field/frequency and based on the maximum value of the FWHM vs angle, which is on the order of 4\,kHz at 7\,T. Our calculated powder pattern agrees quite well with the spectrally dominant resonance in the 2~\% Li-intercalated Ni$_2$P$_2$S$_6$. The second peak, attributed to P sites that are sensitive to Li intercalation, is absent.

\bibliography{Ni2P2S6_NMR}

\end{document}